\documentclass[journal]{IEEEtran}
\usepackage{array}
\usepackage{graphicx}
\usepackage{caption}
\usepackage{subcaption}
\usepackage{mathrsfs}
\usepackage{cleveref}

\usepackage[font=small,labelfont=bf]{caption}
\usepackage{paralist}

\ifCLASSINFOpdf
\else
\fi
\usepackage{amsmath,amsfonts,amsthm} 
\usepackage{paralist}
\usepackage{epstopdf}
\usepackage{bbm}
\newtheorem{theorem}{Theorem}
\newtheorem{lemma}{Lemma}

\newtheorem{definition}{Definition}

\newcommand{\stn}{x_{n}}
\newcommand{\st}{x}
\newcommand{\stnp}{x_{n+1}}
\newcommand{\obs}{y}
\newcommand{\inlk}{\alpha}

\newcommand{\varmi}{H^{-}_{n}}
\newcommand{\varp}{H^{+}_{n}}
\newcommand{\gn}{K_{n}}
\newcommand{\rn}{R_{n}}

\newcommand{\vertexnum}{M}
\newcommand{\network}{G}
\newcommand{\Vertexset}{V}
\newcommand{\edgeset}{E}
\newcommand{\vertexset}{\{1,\hdots,M\}}
\newcommand{\degdist}{\rho}
\newcommand{\nodem}{m}
\newcommand{\degreediff}{D}
\newcommand{\tp}{P}
\newcommand{\atp}{\bar{P}}
\newcommand{\prob}{\mathbb{P}}
\newcommand{\finaltime}{N}
\newcommand{\stv}{\stn}
\newcommand{\bst}{\bar{\st}}
\newcommand{\numagents}{M}
\newcommand{\dev}{\tilde{\st}}
\newcommand{\popn}{x_{k}}
\newcommand{\bpopn}{\bar{x}_{k}}
\newcommand{\mtgk}{v_{k}}
\newcommand{\p}{\prime}

\newcommand{\decay}{\delta}

\newcommand{\nndeg}{m_{d,k}}
\newcommand{\degtr}{H}
\newcommand{\difth}{\lambda_{*}}

\newcommand{\dtime}{n}
\newcommand{\sample}{\gamma}
\newcommand{\degg}{l}
\newcommand{\seq}{l}

\newcommand{\beq}{\begin{equation}}
\newcommand{\eeq}{\end{equation}}
\newcommand{\nodeobs}{\hat{\bst}}
\newcommand{\lbr}{\{}
\newcommand{\rbr}{\}}
\newcommand{\steady}{\pi}
\newcommand{\weight}{W}
\newcommand{\dtme}{k}

\newcommand{\odz}{z_k}

\newcommand{\Obdz}{B_{\odz}}
\newcommand{\hdegdist}{\hat{\degdist}}
\newcommand{\dtmep}{k+1}

\title{Tracking Infection Diffusion in Social Networks: Filtering Algorithms and Threshold Bounds}

\begin{document}
\author{\hspace{-2cm} {Vikram Krishnamurthy\IEEEauthorrefmark{1}, {\em Fellow, IEEE}} \quad \quad 
Sujay Bhatt\IEEEauthorrefmark{2}  \quad \quad
Tavis Pedersen\IEEEauthorrefmark{3} \\
 
\thanks{\IEEEauthorrefmark{1} and \IEEEauthorrefmark{2} are with the Department of Electrical and Computer Engineering,
Cornell University, Ithaca, NY 14850. \IEEEauthorrefmark{3} is with the Department of Electrical and Computer Engineering at the University of British Columbia, Vancouver, Canada, V6T1Z4.

Email: {\tt \IEEEauthorrefmark{1}vikramk@cornell.edu},
{\tt \IEEEauthorrefmark{2}sh2376@cornell.edu} and
{\tt \IEEEauthorrefmark{3}tpedersen@ece.ubc.ca}}}


\maketitle
\begin{abstract}
This paper deals with the statistical signal processing over graphs for tracking infection diffusion in social networks. Infection (or Information) diffusion is modeled using the Susceptible-Infected-Susceptible (SIS) model. Mean field approximation is employed to approximate the discrete valued infected degree distribution evolution by a deterministic
ordinary differential equation for obtaining a generative model for the infection diffusion. The infected degree distribution is shown to follow polynomial dynamics and is estimated using an exact non-linear Bayesian filter. We compute posterior Cram\'er-Rao bounds to obtain the fundamental limits of the filter which depend on the structure of the network. Considering the time-varying nature of the real world networks, the relationship between the diffusion thresholds and the degree distribution is investigated using generative models for real world networks.  In addition, we validate the efficacy of our method with the diffusion data from a real-world online social system, Twitter. We find that SIS model is a good fit for the information diffusion and the non-linear filter effectively tracks the information diffusion. 
\end{abstract}


\section{Introduction}
Statistical signal processing on graphs is an emerging field in which the structural properties of the graph are utilized to derive statistical inference algorithms. As described in~\cite{Pin08}, there is a wide range of social phenomena such as diffusion of technological innovations, cultural fads, and economic conventions~\cite{Cha04} where individual decisions are influenced by the decisions of others. In this paper, we consider social networks represented as graphs and we are interested in analyzing the manner in which information (or infection) spreads through the network.   A large body of research on social networks has been devoted to the diffusion of information (e.g., ideas, behaviors, trends) \cite{Gra78}, and particularly on finding a set of target nodes so as to maximize the spread of a given product \cite{MR07,Che09}. 

\subsection*{Organization and Main Results}

Sec.~\ref{sec:modeldef} presents the \textit{Susceptible Infected Susceptible} (SIS) model for infection diffusion in the network. The key result is that the mean field dynamics approximates the discrete-valued  infected degree distribution evolution by a deterministic ordinary differential equation.  The mean field dynamics yield a tractable model for Bayesian filtering in order to estimate the infected degree distribution given a sampling procedure for the social network. Although, the formulation and mean field dynamics are  known, the proof presented has tutorial value and uses a martingale-based Azuma-Hoeffding inequality. From a signal processing point of view, the mean field dynamics has an interesting interpretation: it resembles a stochastic approximation algorithm; however, in our case, it constitutes an underlying generative model instead of an algorithm.

The mean field dynamics of Sec.~\ref{sec:modeldef} yields a dynamical system whose state (infected degree distribution of network) evolves with polynomial dynamics. Sec.~\ref{sec:nl_fil}  uses a recent result in Bayesian filtering to obtain an exact (finite dimensional filter) for the the infected degree distribution given noisy observations. We examine via numerical examples and posterior Cram\'er-Rao lower bounds, how state estimation over large networks is affected by the network. Numerical examples we demonstrate the difference in performance between power law (Scale Free) and Erd\H{o}s-R\'enyi graphs.

The classical SIS model assumes a fixed underlying social network. Sec.~\ref{sec:df_ev} analyzes the diffusion threshold of a SIS model when the social network evolves over time. Since information diffusion occurs at a faster time scale compared to forming connections in social networks, we consider a two time scale formulation: the degree distribution of the underlying network changes on a slow time scale and the infection diffuses over a faster time scale. Our results generalize the results in \cite{Pin08}, where the underlying network was assumed to be fixed. By using results in stochastic dominance we show that in a preferential attachment model for a randomly evolving graph, the infection diffusion threshold decreases with the attachment probability. To the best of our knowledge, this is new.

Sec.~\ref{sec:dt_tw} illustrates the SIS model and the performance of the Bayesian filter on simulated data and examines the sensitivity of the filter to the underlying graph model (Erd\H{o}s-R\'enyi vs Scale Free). Sec.~\ref{sec:dt_tw} also presents a detailed example using a real Twitter dataset. It is shown via a goodness of fit that SIS is a reasonable model
for information propagation in the  Twitter dataset and that the infected degree distribution can be tracked satisfactorily over time via the Bayesian filter.

\subsection*{Literature}
Susceptible-Infected-Susceptible (SIS) models  for  diffusion of  information in social networks have been extensively studied in~\cite{Pin08,Jac08,Pin06,PV01,Veg07,ZM14,WZT15,VGM14} to  model, for example, the adoption of a new technology in a consumer market.  The literature in SIS and related models for infection diffusion is vast.  Our contribution in this paper is focussed on estimating (tracking) the evolving infected distribution using the mean field dynamics (MFD) as a generative model. The mean field dynamics yield a state space model with polynomial dynamics
and sampling the network yields noisy measurements of the infected degree distribution. 

In this paper, we assume that measurements of the infected degree distribution are obtained by sampling the network. There are various network sampling methodologies studied in the literature, see \cite{VGM14,Gra76,CSW05,Fra05,ANK14,CVSK15,Hec97,Hec02,LF06}. We consider two popular sampling methods: Uniform sampling and Respondent Driven Sampling (RDS) \cite{Hec97,Hec02}. It is shown that under reasonable conditions, by the central limit theorem, that the estimate of the probability that a node is infected in a large population (given its degree) is Gaussian under these sampling methods. Filtering algorithms for polynomial dynamical systems in Gaussian noise were recently developed in \cite{HB14}. It was shown in \cite{HB14} that one can devise a finite dimensional filter (based on the Kalman filter) to compute the conditional mean estimate. Therefore, for MFD, it is possible to obtain an exact Bayesian filtering algorithm for tracking the infected degree distribution. 

To determine a lower bound for the mean square error of the optimal filter, we compute the posterior Cram\'er-Rao bounds \cite{TMN98}. It is observed that the performance of the optimal filter is relatively insensitive to the underlying network degree distribution. 

The main result in Sec.~\ref{sec:df_ev} deals with analyzing how the SIS model is affected when the underlying social network evolves with time. On networks having fixed degree distribution, \cite{Pin08} identified conditions under which a network is susceptible to an epidemic using a mean-field approach and provided a closed form solution for the infection diffusion threshold. The diffusion properties of networks was investigated using stochastic dominance of their underlying degree distributions like in \cite{Jac07,JR07}. However, as real world networks are time evolving, we extend the analysis of diffusion thresholds to time evolving networks using generative models for the underlying network evolution. \cite{HVY14} provides a stochastic approximation algorithm and analysis on a Hilbert space for tracking the degree distribution of evolving random networks with a duplication-deletion model. There are various generative models for large real world networks in the literature, see \cite{CL06}, \cite{LCKF05}, \cite{HVY14}, and the references therein. In this paper, we consider one such model, namely, the preferential attachment model \cite{CL06}, and analyze the connection between the diffusion threshold and the parameters that dictate evolution. The primary motivation for choosing a preferential attachment graph is that it is the simplest graph whose steady state distribution obeys a power law \cite{CL06}, that commonly arises in most real world networks, see~\cite{BA99,BAJ00,AH00}. 

Finally, in Sec.~\ref{sec:dt_tw}, we construct and evaluate the SIS model on a Twitter dataset to track the diffusion of information over time using the filtering algorithms developed in Sec~\ref{sec:nl_fil}. SIS model and its application to the diffusion of information on Twitter has been studied in \cite{KLPM10}, \cite{JDSCR13}. While these papers analyze the effectiveness of SIS on Twitter and other social media, in this paper we not only show via goodness of fit that SIS is a reasonable model, but also show that a filter can satisfactorily track the infection diffusion modeled using mean field dynamics.
\section{SIS Model and Mean Field Dynamics}
\label{sec:modeldef}
Consider a social network where the states of individual nodes evolve over time as a probabilistic function of the states of their neighbors. This evolution can be seen as distributed information processing by the individual nodes to estimate an underlying state process. We model this evolution as infection diffusion over a fast time scale using a SIS model, where the information about the underlying state is accessed using network sampling procedures described in Sec.~\ref{subsec:samp}. \subsection{SIS Model
and Mean Field Dynamics} \label{subsec:inf_diff}
This section discusses the  population model and mean field dynamics for the diffusion of information in a social network. The formulation closely follows  \cite{Pin08}. Consider a social network represented by graph $\network$:
\begin{equation}
\network = (\Vertexset,\edgeset),  \text{ where } \Vertexset= \vertexset, \text{ and } \edgeset  \subseteq \Vertexset \times \Vertexset.
\end{equation}
Here, $\Vertexset$ denotes the set of $\vertexnum$ vertices (users) and $\edgeset $ denotes the set of edges (relationships).
 The degree of a node $m$ is its number of neighbors: $$D^{(m)} = |\{n \in \Vertexset: m,n \in \edgeset\}|,$$ where $|\cdot|$ denotes cardinality.
 
Let $\vertexnum(l)$ denote the number of nodes in the social network $\network$ with degree $l$, and let the degree distribution $ \degdist(l)$ specify the fraction of nodes with degree $l$. That is, for $l=0,1,\ldots,L$,
$$
M(l) =  \sum_{\nodem \in \Vertexset}  I\bigl( \degreediff^{(\nodem)} = l\bigr) , \quad
\degdist(l) =  \frac{ \vertexnum(l) }{\vertexnum} .
 $$
Here, $I(\cdot)$ denotes the indicator function. Since
 $\sum_l \degdist(l)=1$,  $\degdist(l)$ can be viewed as the probability that a node selected  randomly on $\Vertexset$ has connectivity $l$.\\
Assume each node $m$ has two possible states,
$$s_n^{(m)} \in \{ 1 \text{ (infected) },
 2 \text{ (susceptible) } \}. $$
 Let   $\stn(l)$  denote the fraction of nodes with degree $l$  at  time $n$ that are  infected.
We call the $L$ dimensional vector  $\stv$ as the {\em infected population state} at time $n$. So
\begin{equation}
\stn(l) =  \frac{1}{\vertexnum(l)} \sum_{m} I\bigl( D^{(m)} = l, s^{(m)}_n = 1\bigr) , \quad l =1,\ldots, L.
\end{equation}
\subsubsection{SIS Model}
The SIS model assumes that the infected population evolves as follows:
If node $m$ has degree $D^{(m)}= l$, then it jumps from state $i$ to $j$ with probabilities
\begin{multline}
 \tp_{ij}(l,a) = \prob\left(s^{(m)}_{n+1} = j | s^{(m)}_n = i, D^{(m)}=l, F_n^{(m)} = a \right)  \\
 i,j \in \{1,2\}.
\label{eq:tpsocial} 
\end{multline}
Here, $F_n^{(m)}$ denotes the number of infected neighbors of node $m$ at time $n$.
In words, SIS model assumes that the transition probability of a node depends on its degree  and the number of infected  neighbors.\\
The infected population state is updated as
\begin{multline} {\label{eq:st_up}}
\stnp(l) =  \stn(l) +  \frac{1}{\vertexnum(l)} [ I(s_{n+1}^{(l)} = 1, s_n^{(l)} = 2)  \quad \\
-  I (s_{n+1}^{(l)} = 2, s_n^{(l)} = 1) ] 
 \end{multline}
According to (\ref{eq:st_up}), the infected population state changes by $\frac{1}{\vertexnum(l)}$ at every timestep depending on the transition probabilities.
  
The following statistic forms a convenient parametrization of the transition probabilities of $\stv$. Define $\inlk(\stn)$ as the probability that  a uniformly sampled   link  in the network at time $n$ has at least one node that is infected. We call $\inlk(\stn)$  as the
{\em infected
link probability}. Clearly,
\begin{align}\label{eq:linkprob}
\inlk(\stn)
&= \frac{ \sum_{l=1}^{L}\text{(\# of links from infected node of degree $l$)}}
{\sum_{l=1}^{L}\text{(\# of links  of degree $l$)}}  \\
&=
 \frac{ \sum_{l=1}^{L} l \, \degdist(l)\, \stn(l) } {\sum_{l}^{L} l \, \degdist(l) }.  \nonumber
\end{align}
Let $\prob(a|l) = \prob(\text{$a $ out of $l$ neighbors infected})$. In terms of the infected link probability $\inlk$, we can now specify  the scaled transition probabilities\footnote{The transition probabilities are scaled by the degree distribution $\degdist{l}$ for notational convenience. Indeed, since $\vertexnum(l) = \vertexnum \degdist(l)$,
by using these scaled probabilities we can express the dynamics of the process $\st$ in terms of the same-step size $1/\vertexnum$.
We assume that  the degree distribution $\degdist(l)$, $l \in \{1,2,\ldots,L\}$, is uniformly bounded away from zero.} of the process
$\stv$:
\begin{align}
\atp_{21}(l,\inlk_n) &{\overset{\Delta}=}
 \frac{1}{\degdist(l)} \prob\left(\stnp(l) =  \stn(l)+ \frac{1}{\vertexnum(l) } \right) \nonumber  \\
&= (1-\stn(l))  \,  \sum_{a = 0}^l \lambda \tp_{21}(l,a) \,  \prob
(a|l) \nonumber
\\ \label{Eq:atp12def}
& = (1-\stn(l))  \,  \sum_{a = 0}^l \lambda \tp_{21}(l,a) \binom{l}{a}
\inlk_n^a  (1 - \inlk_n)^{l-a},\\
\atp_{12}(l,\inlk_n)  &{\overset{\Delta}=} \frac{1}{\degdist(l)}\,
\prob \left(\stnp(l) =  \stn(l) - \frac{1}{\vertexnum(l) } \right) \nonumber \\
&=
\stn(l) \sum_{a = 0}^l \lambda  \tp_{12}(l,a) \binom{l}{a} \inlk_n^a  (1 - \inlk_n)^{l-a}. \label{Eq:atp21def}
\end{align}
where $\tp_{12},\tp_{21}$ are defined in (\ref{eq:tpsocial}).
In (\ref{Eq:atp12def}) and (\ref{Eq:atp21def}), the notation  $\inlk_n$ is the short form for $\inlk(\stn)$ and $\lambda>0$ is a scaling factor referred to as the diffusion parameter. 
\subsubsection{Mean Field Dynamics} 
We are interested in modeling the evolution of infected population using mean field dynamics. An importatnt feature of the mean field dynamics model is that it has a state of dimension $L$ compared to the intractable state dimension $\Pi_{l=1}^{L} M(l)$ of the infected degree distribution vector $\bst$.
Denote the unit $L-1$ dimensional simplex as $\Pi(L)$:
\begin{equation}
{\hspace{-0.6cm}}\Pi(L){\overset{\Delta}{=}} \{ \st \in\mathbb{R}^{L} : \textbf{1}_{L}'\st = 1, 0\leq \st(i) \leq 1 ~ \text{for all} ~ i\in \{ 1,2, \hdots, L \} \}
\end{equation}
Note that the infected degree vector $\bst_n$ is not a probability distribution as it contains the relative density of infected nodes given the degree. However, scaling each node of degree $l$ by $\frac{1}{\degdist(l)}$, it is seen that the infected degree distribution $\bst_n \in \Pi(L)$.  

\begin{theorem}[Mean Field Dynamics, \cite{Kri16}] \label{thm:mfd}
Consider the deterministic mean field dynamics process with state $\bst_n \in \Pi(L)$ (the $L-1$ dimensional unit simplex):
\begin{equation}  \label{eq:mfd}
 \bst_{n+1}(l)   =  \bst_{n}(l) +  \frac{1}{\vertexnum} \left[ \atp_{12}(l,\inlk(\bst_n)) -
   \atp_{21}(l,\inlk(\bst_n)\right]  
   \end{equation}  
Consider the martingale representation of the Markov chain $\st_n \in \Pi(L)$ (the $L-1$ dimensional unit simplex):
\begin{equation} \label{eq:mrmk}
 \st_{n+1}  = \st_n  + \frac{1}{\numagents}  \left[\tp_{12}(\st)-\tp_{21}(\st)\right] + v_n, \quad \bst_0 = \st_0.  
\end{equation} 
where $v_n$ is a martingale difference process with $\|v_n\|_2 \leq \frac{\Gamma}{\numagents}$ for some positive constant $\Gamma$.
Then for a time horizon of $\finaltime$ points, the deviation between the mean field dynamics $\bst_n$ in (\ref{eq:mfd}) and actual population distribution $\stn$ in (\ref{eq:mrmk}) satisfies
\begin{equation*}
 \label{eq:mfdev}
\prob \bigl\{ \max_{0 \leq n\leq \finaltime} \left\| \bst_n - \stn \right\|_\infty \geq \epsilon\bigr\}
\leq  C_1  \exp(-C_2 \epsilon^2 \numagents)
\end{equation*}
for some positive constants $C_1$ and $C_2$ providing $\finaltime = O(\numagents)$.
\end{theorem}
 The proof of Theorem~\ref{thm:mfd} is given in the appendix in a simple tutorial form that uses the elementary Azuma Hoeffding inequality. Theorem \ref{thm:mfd} says that for a time horizon of $\finaltime$ points,  the deviation between the mean field dynamics $\bst_{n}$ in (\ref{eq:mfd}) and actual infected distribution $\stn$ in (\ref{eq:mrmk})
  satisfies an exponential bound.

For the purposes of this paper, the key outcome of Theorem~\ref{thm:mfd} is that the mean field system $\bst_n$ has polynomial dynamics. These polynomial dynamics will be exploited in Sec.~\ref{sec:nl_fil} for tracking the infected degree distribution.

 \subsection{Sampling}{\label{subsec:samp}}
 
We now consider the second component of the model, namely, observations obtained by sampling the social network. For social networks with large numbers of nodes, it is often impossible to query each node\footnote{An an example, in the case of the MSM social network there is no comprehensive list of all the members of the social network and the members only respond to surveys when prompted by someone already in their network.}. This necessitates choosing a sampling methodology in order to estimate the infected degree distribution $\bst$. For the purpose of estimating the infected degree distribution $\bst$, the degree distribution $\degdist$ of the entire network is assumed to be known{\footnote{In Sec.~\ref{sec:df_ev}, a Bayesian filter using which the degree distribution $\degdist$ itself can be estimated is outlined.}}. Each sampled node is asked if it is infected or not and the reply (measurement) noted. Below, we consider two popular methods for sampling large networks: 
\subsubsection{Uniform Sampling}
At each period $\dtime$,   $\sample(\degg)$ individuals are
sampled\footnote{For large population sizes $\vertexnum$, sampling with and without replacement are equivalent.}  independently and uniformly  from
the population $\vertexnum(\degg)$ comprising of agents with connectivity  degree $\degg$.
That is, a uniform distributed i.i.d. sequence of nodes, denoted by$\{\nodem_{\seq}, \seq=1: \sample(\degg)\}$, is generated from the population  $\vertexnum(\degg)$.
From these independent samples, the empirical infection distribution $\nodeobs_{\dtime}(\degg)$ of degree $\degg$ nodes at each time $\dtime $ is obtained as
\beq \nodeobs_{\dtime}(\degg) = \frac{1}{\sample(\degg) }  \sum_{
\seq=1}^{\sample(\degg)} \lbr s_{\dtime}^{(\nodem_{\seq})}=1\rbr.  \label{eq:sentiment} \eeq
At each time $\dtime$, the empirical  distribution $\nodeobs_{\dtime}$ can be viewed as noisy observations of the infected distribution $\bst_{\dtime}$.

\subsubsection{MCMC Based Respondent-Driven Sampling (RDS)} \label{subsec:RDS}
Respondent-driven sampling~(RDS) was  introduced by Heckathorn~\cite{Hec97,Hec02} as an approach for sampling from hidden populations
in social networks and  has gained enormous popularity in recent years. RDS  is a variant of the well known method of snowball sampling where current sample members recruit future sample members. The RDS procedure is as follows:  A small number of people in the target
population serve as seeds. After participating in the study, the seeds recruit other people they know through the social network in the target population. The sampling continues according to this procedure  with current sample members recruiting
the next wave of sample members until the desired sampling size is reached. Typically,  monetary compensations are provided for participating in the data collection and recruitment.

RDS can be viewed as a form of Markov Chain Monte Carlo~(MCMC) sampling (see~\cite{GS09} for an excellent exposition).
 Let $\{\nodem_{\seq},\seq = 1:\sample(\degg)\}$  be the realization of an aperiodic irreducible Markov chain with
state space  $\vertexnum(\degg)$ comprising of nodes
of degree $\degg$. This Markov chain models the individuals of degree $\degg$ that are snowball sampled, namely, the first individual $\nodem_{1}$ is sampled and then recruits the second
individual $\nodem_{2}$ to be sampled, who then recruits $\nodem_{3}$ and so on.
Instead  of the independent sample estimator~(\ref{eq:sentiment}),
an asymptotically unbiased MCMC estimate is then generated as
\beq \frac{ \sum_{\seq = 1}^{\sample(\degg)}  \frac{I(s_{\dtime}^{(\nodem_{\seq})}=1)}{\steady{(\nodem_{\seq}})} }{ \sum_{\seq=1}^{\sample{(\degg)}}  \frac{1}
{\steady{(\nodem_{\seq}})}}
\label{eq:mcmcrds}
\eeq
where  $\steady(\nodem)$, $\nodem \in \vertexnum(\degg)$, denotes the stationary distribution of the Markov chain.  For example, a reversible Markov chain  with
prescribed stationary distribution  is straightforwardly generated by the Metropolis Hastings algorithm.

In RDS, the transition matrix  and, hence, the stationary distribution $\steady$
in  the estimator~(\ref{eq:mcmcrds})
 is specified as follows: Assume that  edges between any two nodes $i$ and $j$ have symmetric weights $\weight_{ij}$
(i.e.,
 $\weight_{ij} = \weight_{ji}$, equivalently, the network is undirected). In RDS,
node  $i$ recruits node $j$ with transition probability
  $\weight_{ij}/ \sum_{j} \weight_{ij}$. Then, it can be easily seen that
the stationary distribution is
$\pi(i) = \sum_{j \in \Vertexset} \weight_{ij}/ \sum_{i \in \Vertexset, j \in \Vertexset} \weight_{ij}$. Using this stationary
distribution, along with the above transition probabilities for sampling agents in~(\ref{eq:mcmcrds}), yields the RDS algorithm.

It is well known that a Markov chain over a non-bipartite connected undirected network is aperiodic. Then, the initial seed for the RDS
algorithm can be picked arbitrarily, and the above estimator is an asymptotically unbiased estimator.

The key outcome of Sec.~\ref{subsec:samp} is that by the central limit theorem (for an irreducible aperiodic finite state Markov chain), the
estimate of the probability that a node is infected in a large population (given its degree) is Gaussian. Therefore, the sample observations can be expressed as 
\begin{equation} \label{eq:obs_inf}
\obs_{n} = C \bst_n + v_n
\end{equation}
where $v_n \sim \mathscr{N}(\mathbf{0}, \mathbf{R})$ is the observation noise with the covariance matrix $\mathbf{R}$ and observation matrix $C$ dependent on the sampling process and $\bst_n \in \mathbb{R}^{L}$ is the polynomial function of the infected degree probabilities (\ref{eq:mfd}).

\section{Non-linear filter for Bayesian Tracking of Infected Distribution}{\label{sec:nl_fil}}
In Sec.~\ref{sec:modeldef}, we formulated the mean field dynamics for the infected degree distribution as a polynomial dynamical system (\ref{eq:mfd}) with linear Gaussian observations (\ref{eq:obs_inf}) due to sampling the network. In this section, we consider Bayesian filtering of the degree infection probabilities for large networks.
\subsection{Optimal Filter for Polynomial Dynamics}
\label{sec:filter}
To estimate the infected degree  distribution using the sampled observations (\ref{eq:obs_inf}), we employ the optimal Bayesian filter recently developed in \cite{HB14} for polynomial systems. It is shown in \cite{HB14} that for Gaussian systems with polynomial dynamics, one can devise a finite dimensional filter (based on the Kalman filter) to compute the conditional mean estimate. 

Rather than repeating the optimal filtering equations from \cite{HB14}, to save space we present the relevant terms in the model that are used in the filtering equations. 
Let $\hat{\bst}^-_n$ and $\hat{\bst}^+_n$ denote the priori and posteriori expectation of the state vector $\bst_n$ and let $Y_n = \{ y_{0:n}\}$ denote the observation process. Let
\begin{equation}
f(\bst) = A_0 + A_1 \bst + A_2 \bst \bst^\prime + A_3 \bst \bst \bst^\prime + \dots
\label{eq:polynomialfunction}
\end{equation}
denote the polynomial that dictates the evolution in (\ref{eq:mfd}). Here $A_i \in \mathbb{R}^{L\times L \times \dots \times L}$ is an $i+1$ dimensional tensor and 
$$ A_i \bst \bst \dots \bst^\prime = \sum_{j_1, j_2, j_3, \dots, j_i} A_{(:), j_1, j_2, \dots, j_i} \bst_{j_1} \bst_{j_2} \dots \bst_{j_i}
$$ 
The non-linear filter update equations are given as \cite{HB14}:
\begin{equation}
\begin{split}
\hat{\bst}^-_{n} &= \mathbb{E}[f(\bst_{n-1})|Y_{n-1}]\\
\varmi &= \mathbb{E}[(\bst_n -\hat{\bst}_{n})(\bst_n -\hat{\bst}_{n})^\prime | Y_{n-1}] \\
\gn &= \varmi C^\prime(\rn +C \varmi C^\prime)^{-1}\\
\hat{\bst}^+_{n} &= \hat{\bst}^-_{n} +\varmi C^\prime(\rn +C\varmi C^\prime)^{-1}(\obs_{n} -C \hat{\bst}^- _{n}) \\
\varp &= (I-\gn C) \varmi (I - \gn C)^\prime + \gn \rn \gn^{\prime}
\end{split}
\end{equation}
These estimates are computed using the higher order moments of the Gaussian conditional random variable $\bst_n -\hat{\bst}_{n}$: $\mathbb{E}[ (\bst_n -\hat{\bst}_{n})^2 ]) = H_n$,   $\mathbb{E}[ (\bst_n -\hat{\bst}_{n})^6 ]) = 15H^3_n$. The filter relies upon being able to compute the expectation  $\mathbb{E}[f(\bst_{n-1}) f^\prime(\bst_{n-1})|Y_{n-1}]$ in terms of $\hat{\bst}_{n-1}$ and $\varmi$, which is possible when $f(\cdot)$ is a polynomial function.

To derive the filter expressions for the infection dynamics, (\ref{eq:mfd}) is expanded and terms are grouped according to their order in the state $\bst_{n}$ to generate the tensors $A_{i}$ of (\ref{eq:polynomialfunction}). Consider $\bst_n(l)$:
\begin{equation}
\begin{split}
\bst_{n+1}(l)  &=  \bst_{n}(l) +  \frac{1}{\vertexnum} \left[ \atp_{12}(l,\inlk_n)) -
\atp_{21}(l,\inlk_n) \right]  \\
&= \bst(l) \sum_{a = 0}^l \lambda \tp_{12}(l,a) \binom{l}{a} \inlk_n^a (1 - \inlk_n)^{l-a} \\ &\quad +   (1-\bst(l))  \,  \sum_{a = 0}^l \lambda \tp_{21}(l,a) \binom{l}{a}
\inlk_n^a  (1 - \inlk_n)^{l-a} \label{eq:stateevo}
\end{split}
\end{equation}
The average degree is $\sum_{l}^{L} l \, \degdist(l)$ and the link probability given in (\ref{eq:linkprob}) can be expressed as $\inlk_n = \phi^\prime \bst_n$, where $\phi$ is defined as 
\begin{equation*}
\phi = \left[ \frac{\degdist(1)}{\sum_{l}^{L} l \, \degdist(l)} , \frac{2 \degdist(2)}{\sum_{l}^{L} l \, \degdist(l)}, \dots,  \frac{L \degdist(L)}{\sum_{l}^{L} l \, \degdist(l)} \right]^\prime  
\end{equation*}
We expand  (\ref{eq:stateevo}) into a sum of terms that are polynomial in $\bst_{n}$. We illustrate this expansion for degree $l=2$, noting that expansions of any other degree follow the same process. For all terms with factor $P_{12}$ there is a corresponding term with factor $P_{21}$, so for convenience we will account for all of the former with $\Omega$ and the latter with $\Omega^*$. $\Omega$ is then all the terms in an expanded (\ref{eq:stateevo}), with $l=2$, containing $\tp_{12}$. 
\begin{multline}
\Omega  =   \lambda \left[ \frac{1}{\vertexnum}  \tp_{12}(2,0) (\phi^\prime \bst_{n}) ^2 + \frac{2}{\vertexnum} \tp_{12}(2,1)(\phi^\prime \bst_{n}) \right. \\ - \frac{2}{\vertexnum} \tp_{12}(2,1) (\phi^\prime \bst_{n}) ^2 + \frac{1}{\vertexnum} \tp_{12}(2,2) \\ - \frac{2}{\vertexnum} \tp_{12}(2,2) (\phi^\prime \bst_{n}) +\frac{1}{\vertexnum} \tp_{12}(2,2) (\phi^\prime \bst_{n}) ^2 \\ - \frac{\bst_{n}}{\vertexnum} \tp_{12}(2,0) (\phi^\prime \bst_{n}) ^2 -\frac{2 \bst_{n} }{\vertexnum} \tp_{12}(2,1) (\phi^\prime \bst_{n}) \\ + \frac{2 \bst_{n}}{\vertexnum} \tp_{12}(2,0) (\phi^\prime \bst_{n}) ^2 - \frac{\bst_{n}}{\vertexnum} \tp_{12}(2,2) \\ + \frac{2 \bst_{n}}{\vertexnum} \tp_{12}(2,2) (\phi^\prime \bst_{n}) - \frac{\bst_{n}}{\vertexnum} \tp_{12}(2,2) (\phi^\prime \bst_{n}) ^2 \left. \vphantom{\frac{1}{\vertexnum} } \right]
\end{multline} 
and so
\begin{equation}
\bst_{n+1}(2)  =   \bst_{n}(2) + \Omega + \Omega^* \label{eq:polyt}
\end{equation} 
By grouping all of the terms of (\ref{eq:polyt}) by their order in $\bst_{n}$ we can generate the tensors of (\ref{eq:polynomialfunction}). The contributions to the tensors of (\ref{eq:polynomialfunction}) by $\Omega$ are:
\begin{equation}
\begin{split}
A_{0}(2) &= \frac{ \lambda \tp_{12}(2,2)}{\vertexnum} \\
A_{1}(2,:) &= \phi \left[  \frac{ 2 \lambda  (\tp_{12}(2,1) - \tp_{12}(2,0))}{\vertexnum} \right]\\
A_{2}(2,:,:) &= \phi  \phi^\prime  \left[   \frac{ \lambda (\tp_{12}(2,0) - 2 \tp_{12}(2,1) +  \tp_{12}(2,2))}{\vertexnum} \right]\\
A_{2}(2,2,:) &= \phi \left[  \frac{ 2 \lambda  (\tp_{12}(2,1) - \tp_{12}(2,0))}{\vertexnum} \right] \\
A_{3}(2,2,:,:) &= - \phi \phi^\prime  \left[   \frac{ \lambda (\tp_{12}(2,0) - 2 \tp_{12}(2,1) +  \tp_{12}(2,2))}{\vertexnum} \right]
\end{split}\label{eq:polyt2}
\end{equation}
By following (\ref{eq:polyt2}) for $\Omega$ and $\Omega^*$ for all $l$, we are able to generate all the coefficients in the tensors of (\ref{eq:polynomialfunction}) from $\tp_{12}$, $\tp_{21}$, and $\rho(l)$. We note that the polynomial that defines the dynamics of the network is of order $L^* +1$, where $L^*$ is the highest degree node with complex dynamics, i.e: $\tp_{21}(l,a) = \tp_{12}(l,a) = \kappa \quad \forall \quad l > L^* ~\text{and}~a$, where $\kappa$ is constant with respect to both $l$ and $a$.\\ 
  
\subsection{Posterior Cram\'er-Rao bounds}
\label{sec:PCRLB}
The lower bound for the mean square error of the optimal filter is evaluated using the well known Posterior Cram\'er-Rao Lower Bound (PCRLB), \cite{TMN98}. Below we formulate these for the polynomial dynamical system (\ref{eq:mfd}) and linear Gaussian observations (\ref{eq:obs_inf}) with a brief the derivation in the appendix. \\ 
Since there is no state noise in (\ref{eq:mfd}), to compute the PCRLB, we perturb the state evolution in (\ref{eq:mfd}) with pairwise independent Gaussian random vectors having covariance matrix $\mathbf{Q} =\epsilon I$, replacing the singular state evolution by a perturbed system $p_\epsilon(\bst_{n+1}|\bst_n)$. We have,
\begin{equation} \label{eq:per_st}
{\hspace{-0.5cm}}-\log p_\epsilon(\bst_{n+1}|\bst_n) =  c + \frac{1}{2}\{ \bst_{n+1} - f_n(\bst_n) \}^\prime \textbf{Q}^{-1} \{\stnp -f_n(\bst_n)\} 
\end{equation}
Let $J(X_{n})$ denote the $((n+1)L \times (n+1)L)$ Fisher information matrix{\footnote{The error covariance $\mathscr{P}$ is,
$$ \mathscr{P} = \mathbb{E}\{(\hat{{X}} - {X}) (\hat{{X}} - {X})^\prime \} \geq J^{-1}$$
where $J$ is the Fisher Information matrix, ${X}$ is the state, $\hat{{X}}$ is the state estimate, and ${Y}$ measured data.
The elements of the Fisher information matrix $J$ are given by:
$$
J_{ij} =\mathbb{E} \left( \frac{ \partial^2 \log p_{x,y}(X,Y)}{\partial X_i \partial X_j} \right)
$$ }} of $X_{n}$ \cite{TMN98}, with $X_{n} = [X_{n-1},\stn]$. The following recursion is used to evaluate $J_{n}$ in \cite{TMN98}. Let $Y_n = [Y_{n-1},y_n]$ denote the observation sequence at time $n$, $\prob(X,Y)=p_n$ denote the joint distribution and $\Delta_{x}^y =  \nabla_x \nabla^\prime_y$ denote the vector differential operator. Then~\cite{TMN98}, 
\begin{multline*}
J_n = \mathbb{E}\{-\Delta_{x_n}^{x_n} \text{log} (p_n)\} -   \quad \\
  \mathbb{E}\{-\Delta_{x_n}^{X_{n-1}} \text{log} (p_n)\} [\mathbb{E}\{-\Delta_{X_{n-1}}^{X_{n-1}} \text{log} (p_n)\}]^{-1} \mathbb{E}\{-\Delta_{{X_{n-1}}}^{x_n} \text{log} (p_n)\}
\end{multline*}
 and corresponds to the inverse of the $L\times L$ right lower block of $\left[ J(X_{n})\right]^{-1}$. The recursion for $J_n$ is given by:
$$
J_{n+1} = D_{\epsilon,n}^{22} - D_{\epsilon,n}^{21} \left( J_n +D_{\epsilon,n}^{11}\right)^{-1} D_{\epsilon,n}^{12}
$$
where\footnote{see Appendix for a brief derivation}
\begin{equation}
\begin{split}
 D_{\epsilon,n}^{11} &=  \frac{1}{\epsilon} \mathbb{E} \{ (\nabla_{\bst_n}f_n^\prime(\bst_n)) (\nabla_{\bst_n}f_n^\prime(\bst_n))^\prime \} \\
D_{\epsilon,n}^{12} &=   \frac{1}{\epsilon}  \mathbb{E} \{ \nabla_{\bst_n}f_n^\prime(\bst_n) \} \\
D_{\epsilon,n}^{21} &=  \{D_n^{12}\}^\prime \\
D_{\epsilon,n}^{22} &=   \frac{1}{\epsilon} I + C \mathbf{R}_n^{-1} C^\prime \\
\end{split}
\end{equation}
where $C$ is the linear observation matrix and $\mathbf{R}$ is the observation noise covariance matrix of (\ref{eq:obs_inf}) and $\nabla_{\bst_n}f_n^\prime(\bst_n)$ is given as:
$$
\nabla_{\bst_n} f_n^\prime(\bst_n) = \left[ \frac{\partial f}{\partial \bst_n(0)}, \frac{\partial f}{\partial \bst_n(1)}, \dots, \frac{\partial f}{\partial \bst_n(D)}, \right]^\prime 
$$
\begin{equation}
\begin{split} \frac{\partial f}{\partial \bst_n(0)} = 0 +  \frac{\partial}{\partial \bst_n(0)} \left[ A_1 \bst_n \right] + \frac{\partial}{\partial \bst_n(0)} \left[ A_2 \bst_n \bst_n^\prime\right]+ \\ \frac{\partial}{\partial \bst_n(0)} \left[ A_3 \bst_n \bst_n \bst_n^\prime \right] \end{split} 
\end{equation}
Thus 
\begin{equation}
\nabla_{\bst_n} f_n^\prime(\bst_n) = A_1 + (A_2 +A_2^\prime)\bst_n + (A_{3_{ijk}} + A_{3_{jki}} +A_{3_{kji}}) \bst_n \bst_n^\prime
\end{equation}
where $A_{3_{ijk}}$ indicates the ordering of indices of the tensor $A_3$ and is analogous to a higher dimensional transpose. 

\subsubsection*{PCRLB: Erd\H{o}s-R\'enyi vs Scale Free}
We used the above method to compute the PCRLB for the mean field dynamics model (\ref{eq:mfd}), with linear Gaussian observations (\ref{eq:obs_inf}). We first consider a scale-free network with degree distribution $ \degdist(l) \propto  l^{-\gamma}$. Such networks arise in online social networks such as Twitter \cite{KLPM10} and in the link network of the World Wide Web \cite{AH00}. The second network we consider is an Erd\H{o}s-R\'enyi network with degree distribution $ \degdist(l) \propto  \frac{e^{-\lambda l }}{l!} $. Fig.5 shows the PCRLB for a scale-free network with $\gamma =2.7$ and an Erd\H{o}s R\'enyi network with $\lambda=2.7$. The value $\lambda=2.7$ was chosen since it is similar to the the out-degree of the World Wide Web, see \cite{BKMRRSTW00}. For both networks, the infection transition probabilities were a random stochastic matrix. The displayed mean square errors are the average of 100 independent simulations. 

\begin{figure}
     \centering
     \includegraphics[scale=0.275]{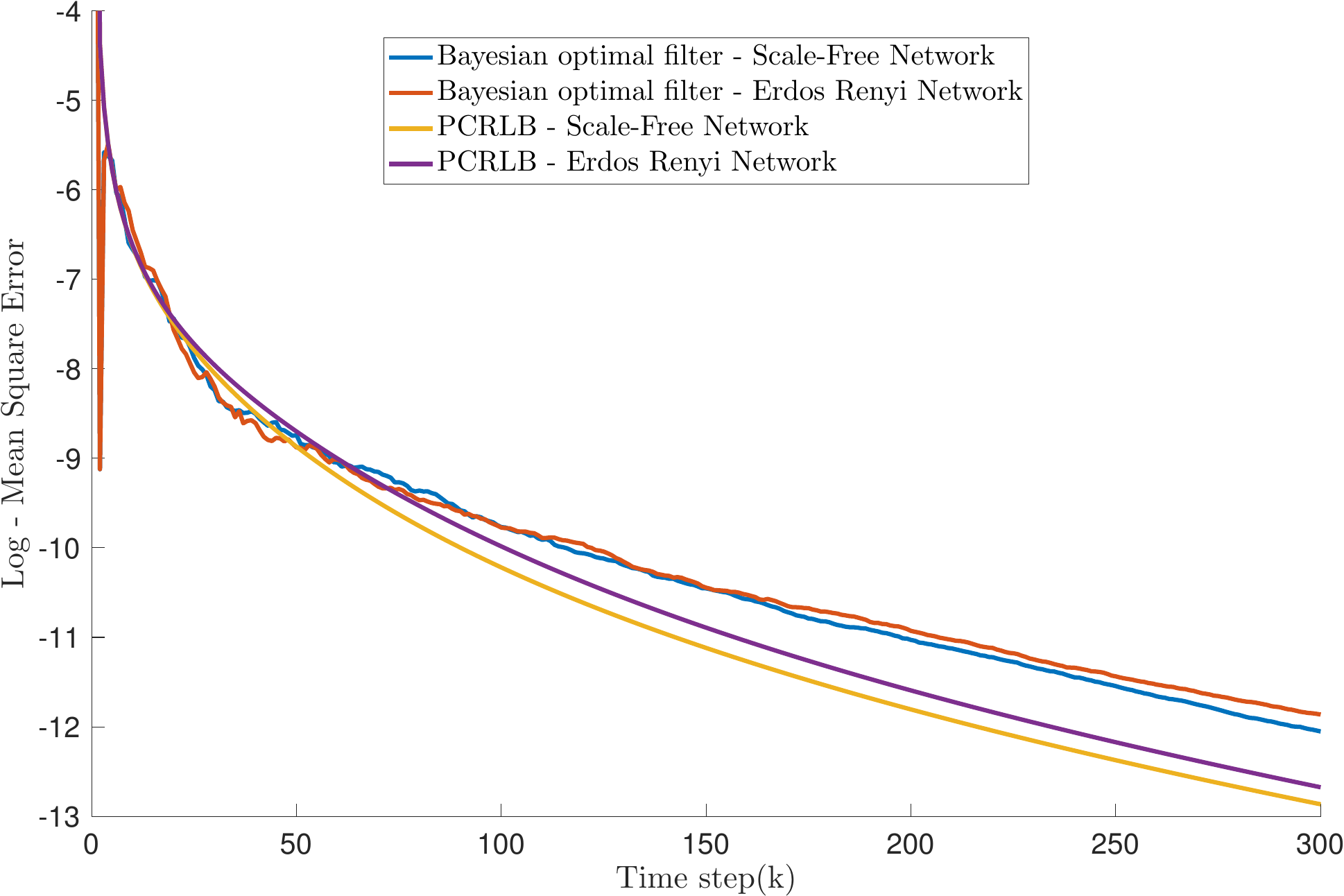}
     \caption{Mean square error and PCRLB for two different networks - Scale-free and Erd\H{o}s-R\'enyi. It can be seen that the filter performs near optimally and has no discernible dependence on the degree distribution. Both PCRLB and its slope are insensitive to the underlying distribution. }
     \label{fig:pc_mse}
 \end{figure}

Interestingly, it can be seen from Fig.~\ref{fig:pc_mse} that both PCRLB and its slope are insensitive to the underlying network structure when observation noise variance, $\mathbf{R}$ in (\ref{eq:obs_inf}), is not network dependent. In Sec.~\ref{sec:dt_tw}, we will show that the performance of the optimal filter is also insensitive to the underlying network structure. These suggest that for tracking the infected degree distribution, precise knowledge of the underlying network distribution is not required.

\section{Analysis of infection diffusion in evolving social networks}\label{sec:df_ev}
So far in this paper, we have discussed estimating infection diffusion in a fixed network. In this section, we consider social networks that evolve with time, represented by changing degree distributions, and analyse their effect on the diffusion of infection over time. Since information diffusion occurs at a faster time scale compared to forming connections in social networks, we consider a two time scale formulation: where the degree distribution of the underlying network changes on a \textit{slow time} scale and the infection diffuses over a \textit{faster time} scale. There are various generative models for large real world networks in the literature, see \cite{CL06}, \cite{LCKF05}, and the references therein. In this paper, we consider the preferential attachment model  discussed extensively  in \cite{CL06}, to model the time evolution of the underlying degree distribution. The primary motivation for choosing a preferential attachment graph is that it is the simplest graph whose steady state distribution obeys a power law \cite{CL06}, which commonly arises in most real world networks, see~\cite{BA99,BAJ00,AH00}.
\subsection{Preferential Attachment Model for Network Evolution} {\label{subsec:pa_md}}
A network evolving according to the preferential attachment model\footnote{The structural results we present here apply to any dynamical graph model that satisfies the stochastic dominance conditions given in Theorem~\ref{thm:cmp_gr}.} is characterized by two parameters - a probability $p$ and an initial graph $G_{0}$. The graph evolves as follows: 
\begin{compactenum}
\item \textit{Vertex-Step}: A new vertex is added and is connected to a vertex of the existing graph chosen independently with probability proportional to its degree.
\item \textit{Edge-Step}: A new edge is added between two vertices of the graph chosen independently with probability proportional to their degrees.
\end{compactenum}
At each time step, with probability $p$, Vertex-step is realized and with probability $1-p$, Edge-step is realized. \\

We are interested in how the diffusion thresholds (see Def.~\ref{def:df_t} below) change when the graph is evolving with different parameters. In order to determine how the diffusion threshold depends on the graph evolution, we first need to specify a generative model for the evolution.

Let $\nndeg$ denote the number of vertices of degree $d$ at time $\dtme$ and 
$$\degdist_{d,\dtme} = \frac{\mathbb{E}(\nndeg)}{\numagents_\dtme}$$ denote the expected fraction (degree distribution) of vertices of degree $d$ at time $\dtme$, where $\numagents_\dtme$ denotes the total number of nodes or agents at time $\dtme$ in the network. A vertex of degree $d$ at time $\dtme$ could have come from two cases, either it was a vertex of degree $d$ at time $\dtme-1$ and had no edge added to it, or it was a vertex of degree $d-1$ at
time $\dtme-1$ and a new edge was put in adjacent to it. The recursion for the degree distribution $\degdist_{d,\dtme}$ can be expressed as \cite{CL06}:
\begin{equation} 
\degdist_{d,\dtme} = \left( 1-\frac{(2-p)d}{2\dtme}\right) \degdist_{d,\dtme-1} + \left(\frac{(d-1)(2-p)}{2\dtme}\right) \degdist_{d-1,\dtme-1}
\end{equation}
Let $\degdist_{\dtme}  = [\degdist_{1,\dtme} , \degdist_{2,\dtme} ,\hdots,\degdist_{N,\dtme},\degdist_{N^{(+)},\dtme}]$ denote the degree distribution at time $\dtme$, with $\degdist_{N^{(+)},\dtme}$ representing all degrees greater than $N$. The grouping of all states into one compound state is for notational convenience and as will be shown below, it is amenable to analysis. The compound state is modeled as an absorbing state as either edges or vertices are added during network evolution and no deletion takes place - once a node is of degree greater than $d$, it will continue to have degree atleast $d$. In matrix form, the recursion in equation (\ref{eq:deg_ev}) can be written as 
\begin{equation}\label{eq:deg_ev}
\degdist_{\dtme} = \degtr^\prime_{\dtme} \degdist_{\dtme-1}
\end{equation} \label{eq:ro_ev}
where the matrix $\degtr_\dtme$  can be expressed as $\degtr_\dtme = I + \epsilon_\dtme F$ and
\[
{\hspace{0cm}} F
=
\begin{bmatrix}
-(2-p)& (2-p) & 0 & \hdots 0 & 0 \\
0 &  -(2(2-p)) & (2(2-p))&  \hdots 0 & 0 \\
\vdots & \ddots \\
0 & 0 & \hdots & 0 &  0 
\end{bmatrix}
\]
where $F$ is a generator matrix for the graph evolution having row sum equal to $0$ and $\epsilon_\dtme=\frac{1}{\dtme}$. Clearly, the matrix $H$ is a stochastic matrix with row sum equal to $1$. In what follows, we will give sufficient conditions to compare the evolution dynamics of two social networks when the underlying graphs are changing according to a preferential attachment scheme. 

On networks having fixed degree distribution (fixed networks), \cite{Pin08} identified conditions under which a network is susceptible to an epidemic using a mean-field approach and provided a closed form solution for the diffusion threshold for infection diffusion\footnote{
\begin{multline*}
\difth = \frac{\sum_{l}^{L} l \, \degdist(l)}{\sum_{l\geq1} \, l^2 \degdist(l) \tp_{ij}(l,1)},~~\text{where}~\tp_{ij}(l,a)~\text{was defined in }(\ref{eq:tpsocial}).
\end{multline*}}. It was shown that under reasonable conditions on the infection probabilities, the diffusion threshold decreases with the mean preserving spread. In this paper, we extend the analysis of diffusion thresholds to evolving networks by providing sufficient conditions on the parameters that dictate the evolution. 
\subsection{Effect on Diffusion Threshold in SIS model}\label{subsec:e_dt}
In this section, we establish a relation between the addition probability $p$ in the preferential attachment model and the diffusion threshold $\difth$ in the SIS model.
\begin{definition}[\cite{Pin08}]~{The Diffusion Threshold ($\difth$) is}\label{def:df_t}
$$\difth = \inf\{\lambda>0: \bst_\infty \in \mathbb{R}^L_{+} \}$$
where $\bst_\infty$ denotes the asymptotic infected degree distribution in (\ref{eq:mfd}).
\end{definition}
 In words, diffusion threshold $\difth \in \mathbb{R}^+$ is a value of $\lambda$ in (\ref{Eq:atp12def}) and (\ref{Eq:atp21def}), such that starting from a small fraction of infected agents in the network, the dynamics converges to a \textit{positive} fraction of infected agents for all $\lambda > \difth$. The existence of asymptotic infected degree distribution $\bst_\infty$ is given by Lemma~\ref{lem:asm_ro}. For a proof, the reader is referred to \cite{Pin08}. 
 
 Let $Q(\inlk) = \frac{1}{l}\sum_{l \geq 1} l \degdist(l) \frac{\atp_{21}(l,\inlk)}{(1-\st(l))+\atp_{21}(l,\inlk)}$, where  $\atp_{21}(l,\inlk)$ and $(1-\st(l))$ are defined in (\ref{Eq:atp12def}). 
\begin{lemma}[\cite{Pin08}] \label{lem:asm_ro}
 $\bst_\infty$ exists iff  $\frac{dQ(0)}{d\inlk} > 1$. 
\end{lemma} 
In words, there exists an asymptotic infected degree distribution if  $Q(\inlk)$ has a slope greater than $45^0$ at the origin. The asymptotic infected link probability $\inlk_\infty$ can be calculated from $\bst_\infty$ using~(\ref{eq:linkprob}). 

The following theorem characterizes of the diffusion threshold of the SIS model as a function of the addition probability $p${\footnote{It should be noted that the probability $p$ itself can be a function of $\inlk_\infty$ as the degree distribution and infected degree distribution are evolving on different time scales.}}, for a preferential attachment graph. 
\begin{theorem} \label{thm:cmp_gr}
Consider a time evolving preferential attachment network with addition probability $p>0$. For any initial degree distribution $\degdist_0$, let $\degdist_k(p)$ denote the degree distribution at time $\dtme$ and $\difth^p(\dtme)$ denote the diffusion threshold for the network with addition probability $p$. Then,
\begin{compactenum} 
\item $\degdist_k(p)$  is first-order stochastically decreasing in $p$ for every $\dtme>1$, where $\dtme$ denotes the slow-time index. 
\item $\difth^p(\dtme)$ is increasing in $p$.
\end{compactenum}
\end{theorem} 
The proof of Theorem~\ref{thm:cmp_gr} is given in the appendix. The first part of Theorem~\ref{thm:cmp_gr} asserts  that $\degdist_k(p_2) >_{sd}\degdist_k(p_1)$, \footnote{ $>_{sd}$  denotes first-order stochastic dominance (see Appendix for definition)} for $p_1>p_2$, i.e, networks that have higher probability of edge addition always have higher degree distributions as the network evolves. The second part of Theorem~\ref{thm:cmp_gr} asserts that the diffusion threshold increases with the probability $p$ of adding new vertices. This can be interpreted as networks with smaller number of nodes of higher degree requiring a larger fraction of infected individuals to have a positive fraction of infected individuals when the dynamics converges.

\textit{Remark}: The key result above is the stochastic dominance structure for the preferential attachment graph - it is of interest in future work to give sufficient conditions on more general types of dynamic graph models. 

\subsection{Filtering for Estimating the degree distribution}
So far in this paper, it was assumed that the degree distribution of the social network is known. In this section, we outline a Bayesian filter to estimate the degree distribution, when it evolves according to the preferential attachment model\footnote{Recall that Sec~\ref{sec:nl_fil} deals with filtering to track the infected degree distribution in a fixed network.}. 

In Sec.~\ref{subsec:e_dt}, it was shown that the social network structure (degree distribution $\degdist$) plays a significant role in the asymptotic infected degree distribution (Lemma~\ref{lem:asm_ro}). However, since the infection diffusion occurs at a faster time scale ($n$) compared to forming connections in social networks (time scale $k$), the asymptotic distribution $\bst_\infty$ can in turn influence the network rearrangement at a future time $k+1$. For the preferential attachment model of Sec.~$\ref{subsec:pa_md}$, this influence can be modeled as the probability $p$ being dependent on $\inlk_{\infty}$ (which depends on~$\bst_\infty$). 

It is interesting to note that the evolution of degree distribution in (\ref{eq:deg_ev}) has the form of a Chapman-Kolmogorov equation for a Markov chain $\eta$ having the state space $\{1,2,\hdots,{N^{+}}\}$. In other words, Chapman-Kolmogorov equation is a generative model for the evolution of the network. We exploit this property to estimate the degree distribution by deriving a representative sample that captures the link between the degree distribution $\degdist$ and the asymptotic degree distribution $\bst_\infty$. This link could be an important factor in determining the way connections are formed in social networks; see for example, \cite{MSC01}; where, similarity between individuals (Homophily) breeds connection. Nodes which are not previously connected, but being infected currently increases the probability of forming a link at a future time instant. 

Let $\dtme$ = $0,1,\hdots$ denote the slow time index. Let $\odz${\footnote{Observation $\odz$ can be a scalar or a vector depending on the application.}} denote the observation at time $\dtme$. Below, we consider the mode of the asymptotic infected degree distribution $\bst_\infty \in \mathbb{R}^L$ as the representative sample $\odz$  at time $\dtme$. The mode of the infected degree distribution gives the degree with the largest fraction of infected individuals and tracking the mode can provide useful information on the nature of infection diffusion over time $\dtme$. Let the initial estimate be $\hdegdist_0$, which denotes the probability distribution{\footnote{Note that here the degree distribution $\hdegdist$ is interpreted as the probability distribution of the mode of asymptotic infected degree distribution $\bst_\infty$.} } of the mode (of asymptotic infected degree distribution) over the set~$\{1,2,\hdots,{N^{+}}\}$.

Given this observation $\odz$ at time $k$, and the dynamics of the degree distribution (\ref{eq:ro_ev}), define the posterior distribution of the degree distribution as 
$$\hdegdist_{\dtmep} = \prob( \degdist_{\dtmep}| z_1,\ldots, z_{\dtmep}) $$
Then it is easily seen that the evolution of the posterior distribution is given as the Hidden Markov Model (HMM) filter~\cite{Kri16}:
\begin{equation} \label{eq:hmm}
\hdegdist_{\dtme+1} = \frac{\Obdz \degtr_\dtme^\prime \hdegdist_{\dtme}}{\textbf{1}^\prime \Obdz \degtr_\dtme^\prime \hdegdist_{\dtme}}
\end{equation}
where $\Obdz = \text{diag}(\prob(\odz | \eta_k=i))~\text{for}~i \in \{1,2,\hdots,{N^{+}}\}$ is a diagonal matrix. \\
It should be noted that the at time $\dtmep$, both the estimate of the mode and the degree distribution $\hdegdist_{\dtmep}$ are obtained. 

To summarize, (\ref{eq:hmm}) together with the filter (Sec.~\ref{sec:filter}) constitutes a two time scale tracking algorithm:  on the slow time scale, the degree distribution of the social network is updated based on sampling according to (\ref{eq:hmm}). On the fast time scale, these estimates are used in the filter (Sec.~\ref{sec:filter}) to track the infected degree distribution.  We refer to \cite{Kus12} for a formal proof of the optimality of this time two-time scale filtering algorithm.

\section{Numerical Examples and Twitter Data} \label{sec:dt_tw}
This section presents two main results. First, computer simulations are used to illustrate the performance of the filtering algorithm for tracking the evolving infected degree distribution. The sensitivity of the filter to the underlying network (Erd\H{o}s-R\'enyi vs Scale Free) is examined.\footnote{Recall Sec.\ref{sec:PCRLB} examined the sensitivity of the posterior Cram\'er-Rao bounds to the underlying network.}  Second, the SIS model is illustrated using a real  Twitter dataset and the infection diffusion is tracked using the non-linear filter in Sec.\ref{sec:nl_fil}. It is shown using a goodness of fit test that the SIS model yields a satisfactory fit to the Twitter data and also that the Bayesian filter performs satisfactorily. 
\subsection{Filtering on Erdos-Renyi vs Scale-Free Networks} \label{subsec:filt_erpl}
 Recall a scale-free (SF) network has a degree distribution $ \degdist(l) \propto  l^{-\gamma}$ while an Erd\H{o}s-R\'enyi (ER) network has a degree distribution $ \degdist(l) \propto  \frac{e^{-\lambda l }}{l!} $. Given the transition probabilities, degree distribution, and initial conditions, we can simulate trajectories of the mean-field dynamics (\ref{eq:mfd}) and generate observations according to (\ref{eq:sentiment}). Below, we illustrate the effect of sampling on filtering.
\subsubsection{Effect of Sampling on Filtering}\label{subsec:sam_erpl}
Observation noise variance $\mathbf{R}$ in (\ref{eq:obs_inf}) depends on the sampling method employed. Through this dependence, if sampling is network dependent, so is the observation noise variance, and transitively the estimate is network dependent. The effect of sampling on filtering is illustrated using Uniform sampling of Sec.~\ref{subsec:samp}. 

\begin{figure}[h!]
	\centering
	\includegraphics[width=3.5in]{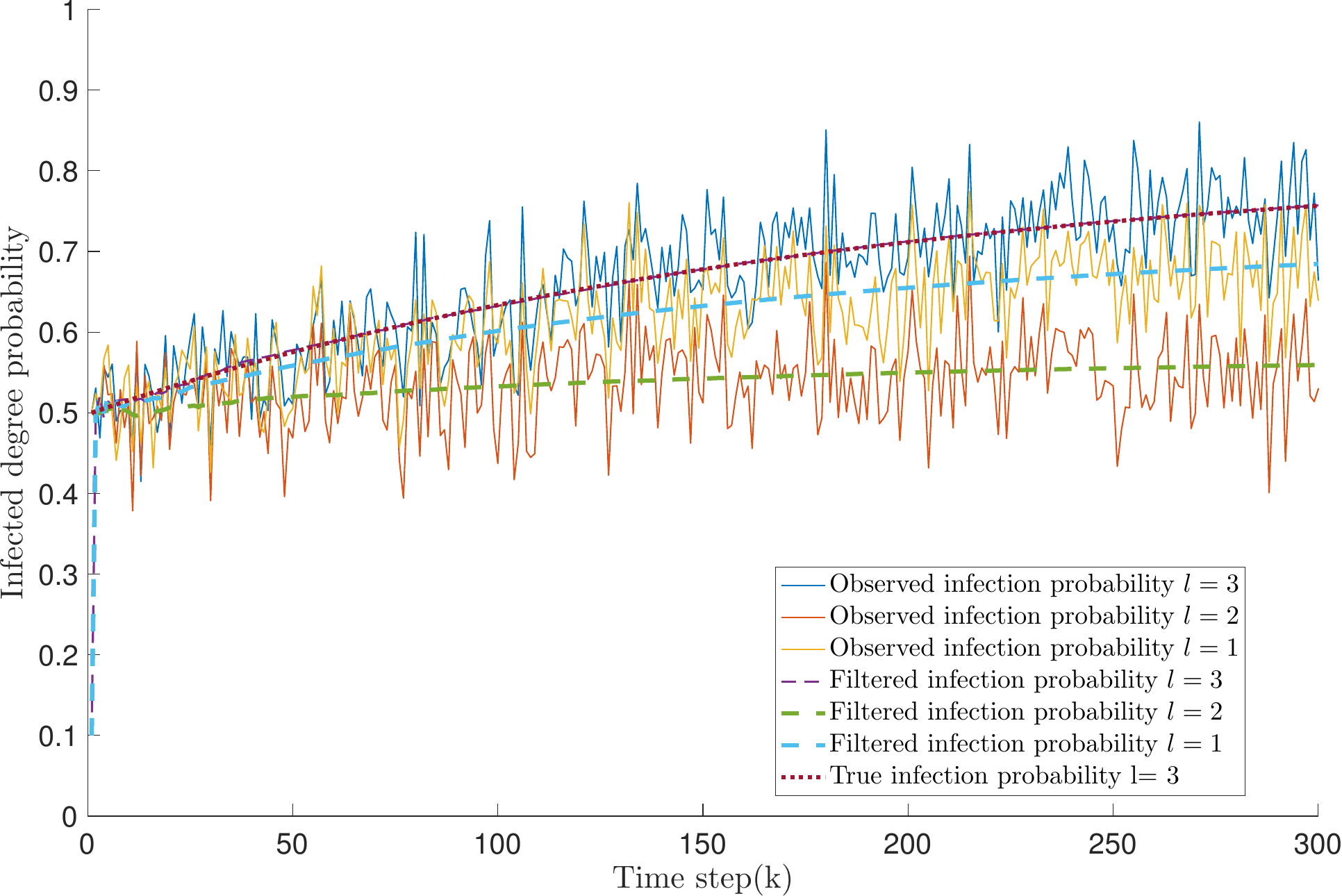}
	\caption{Diffusion of infection probability trajectories and their corresponding filtered states in a scale-free network. It can be seen that the estimates converge to the true state for all degrees. }
	\label{fig:opt_filt}
\end{figure}

\begin{figure}[h!]
	\centering
	\includegraphics[width=0.9\linewidth]{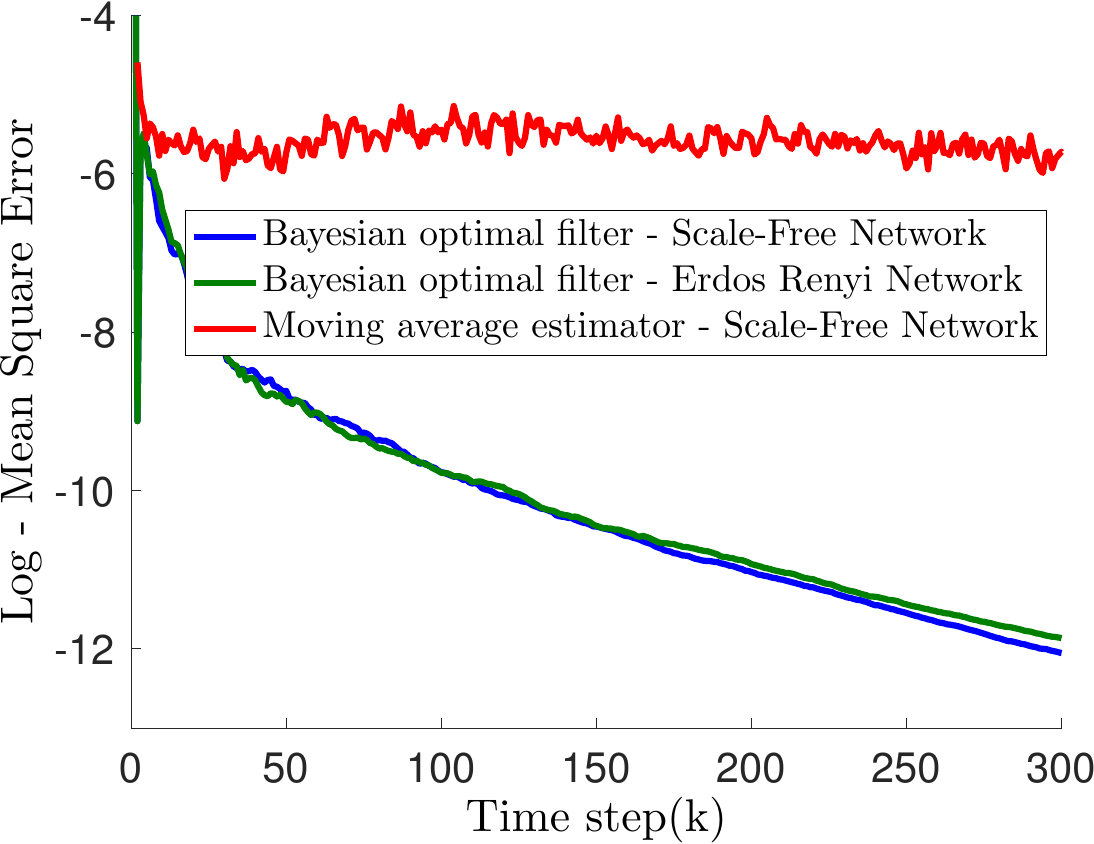}
	\caption{ Mean Squared Error for both Erd\H{o}s-R\'enyi and scale-free networks are compared. For both networks, the same transition probabilities and initial conditions are used. The averages shown have been averaged over 50 simulations. It is observed that for constant observation noise variance, the MSE for both Scale-Free and Erd\H{o}s-R\'enyi networks are indistinguishable. }
	\label{fig:MSE_ERPL}
\end{figure}

\textit{Uniform Sampling}: A simulated diffusion state and observation trajectory are shown in Fig.~\ref{fig:opt_filt}, and the respective mean square filter error is shown in Fig.~\ref{fig:MSE_ERPL}.{\footnote{The performance of the moving average filter is provided for comparison.}}  The transition probabilities are a random stochastic matrix (same for both ER and SF networks); and observations are simulated according to (\ref{eq:obs_inf}) with $C =I$, $\st_{0} =\frac{1}{2} \forall l $, and constant observation noise variance $\mathbf{R} = (5\times10^{-3} )I$, where $I$ is the identity matrix. These parameters were chosen arbitrarily. It is observed that for constant observation noise variance, there is no discernible difference between Scale-Free and Erd\H{o}s-R\'enyi networks in the accuracy of the filtered state estimates. 

\begin{figure}[h!]
	\centering
	\begin{subfigure}[b]{2.6in}
		\includegraphics[width=2.6in]{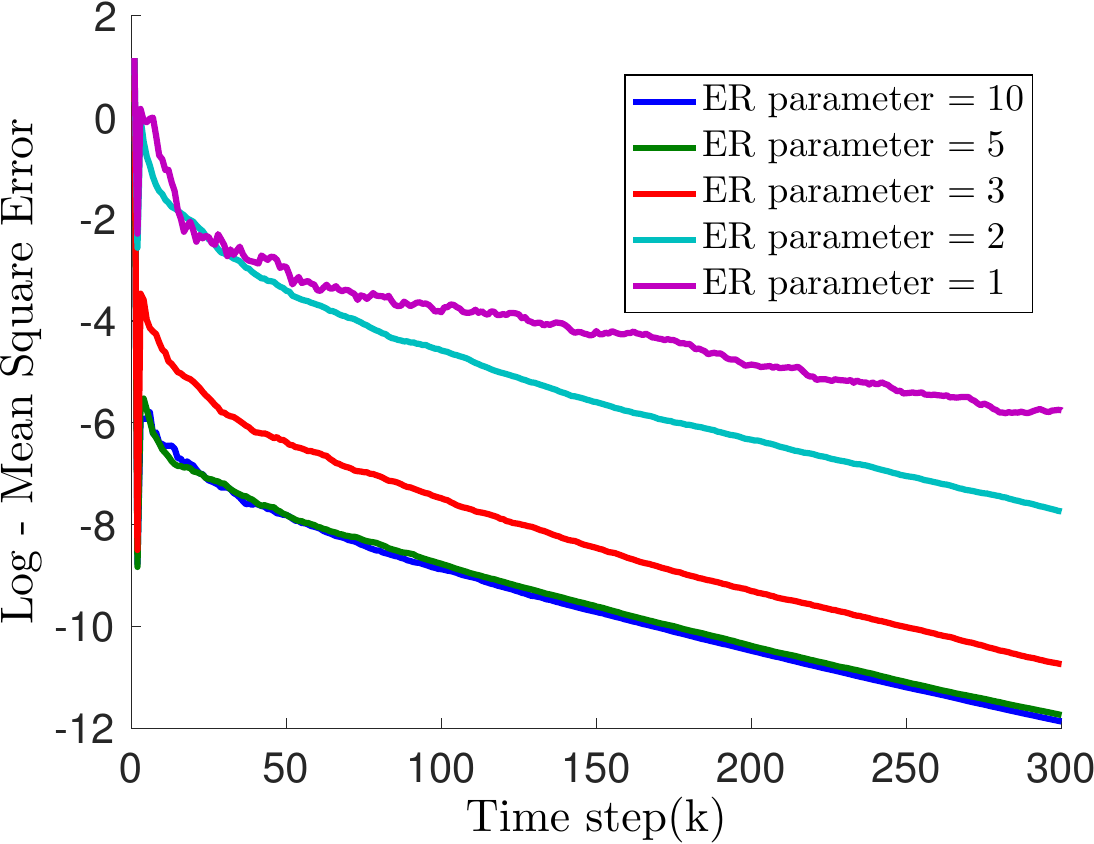}
		\caption{Log mean square error in Erd\H{o}s R\'enyi networks for varying ER parameter, $\lambda$. It can be seen the MSE decreases as $\lambda$ increases. }
		\label{fig:ER_param_vary}
	\end{subfigure}
	~ 
	\begin{subfigure}[b]{2.6in}
		\includegraphics[width=2.6in]{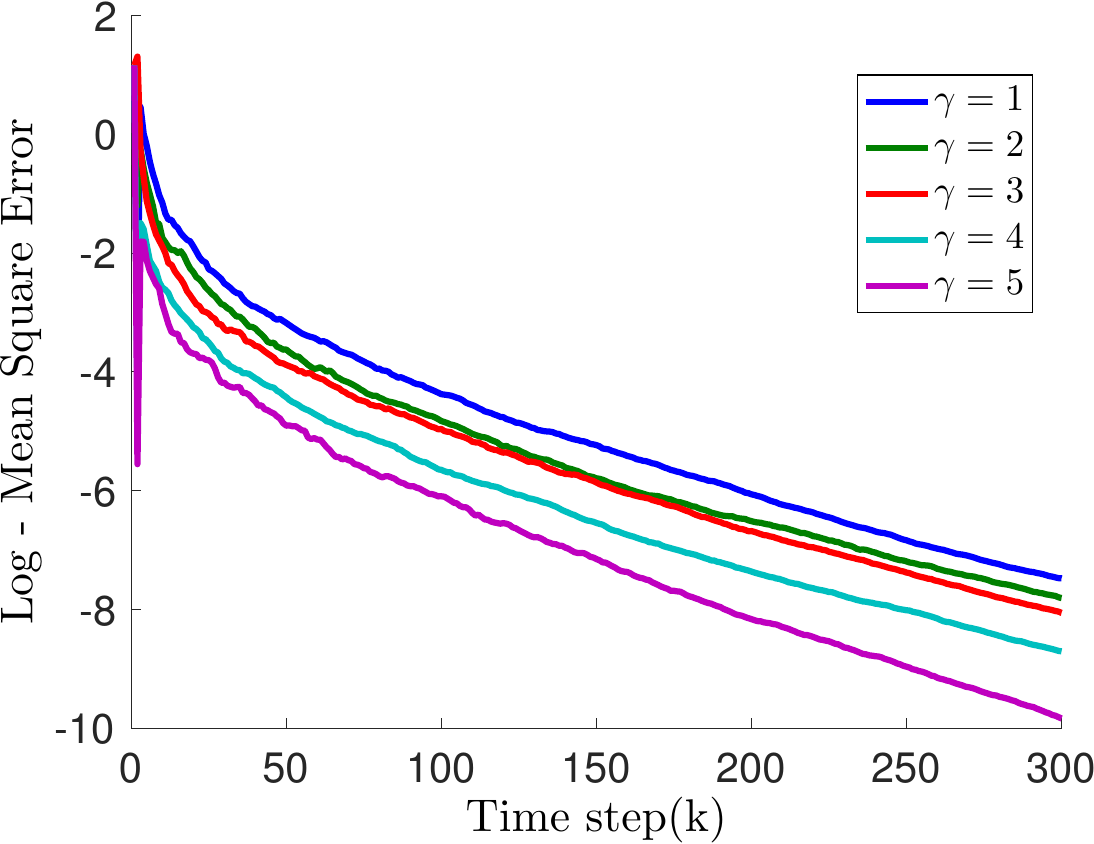}
		\caption{Log mean square error in scale-free networks for varying $\gamma$. For larger scale-free parameter $\gamma$ it is observed that the MSE increases.}
		\label{fig:PL_param_vary}
	\end{subfigure}
	\caption{Mean square error for varying network parameters and network types.}\label{fig:EffectSampling}
\end{figure}

In Fig.~\ref{fig:EffectSampling}, the transition probabilities are a random stochastic matrix; and observations are simulated according to (\ref{eq:obs_inf}) with $C=I$ $\st_{0} =\frac{1}{2} \forall l $ and a non-diagonal \textit{random} observation noise variance matrix $\textbf{R}$. We observe that distributions that are more uniformly spread amongst all degrees result in fewer degrees with high observation noise variance, which in turn reduces the total mean square error. In scale-free networks, as shown in Fig.~\ref{fig:PL_param_vary}, for larger $\gamma$, the degree distribution decays more quickly, thus there is less probability mass on nodes of higher degree and they are less frequently sampled. According to (\ref{eq:sentiment}), fewer samples corresponds to a higher variance and therefore a worse estimate. In Erd\H{o}s R\'enyi networks Fig.~\ref{fig:ER_param_vary}, most of the probability mass is centered around the mean of the degree distribution and so there are fewer degrees with large observation noise variance. It can be seen that ER networks having a higher average degree $\lambda$ have smaller MSE. 

\subsubsection{Sensitivity of Filter Performance to Mis-specified Model} \label{subsec:sen_dd}
\begin{figure}[h!]
	\centering
	\includegraphics[width=0.7\linewidth]{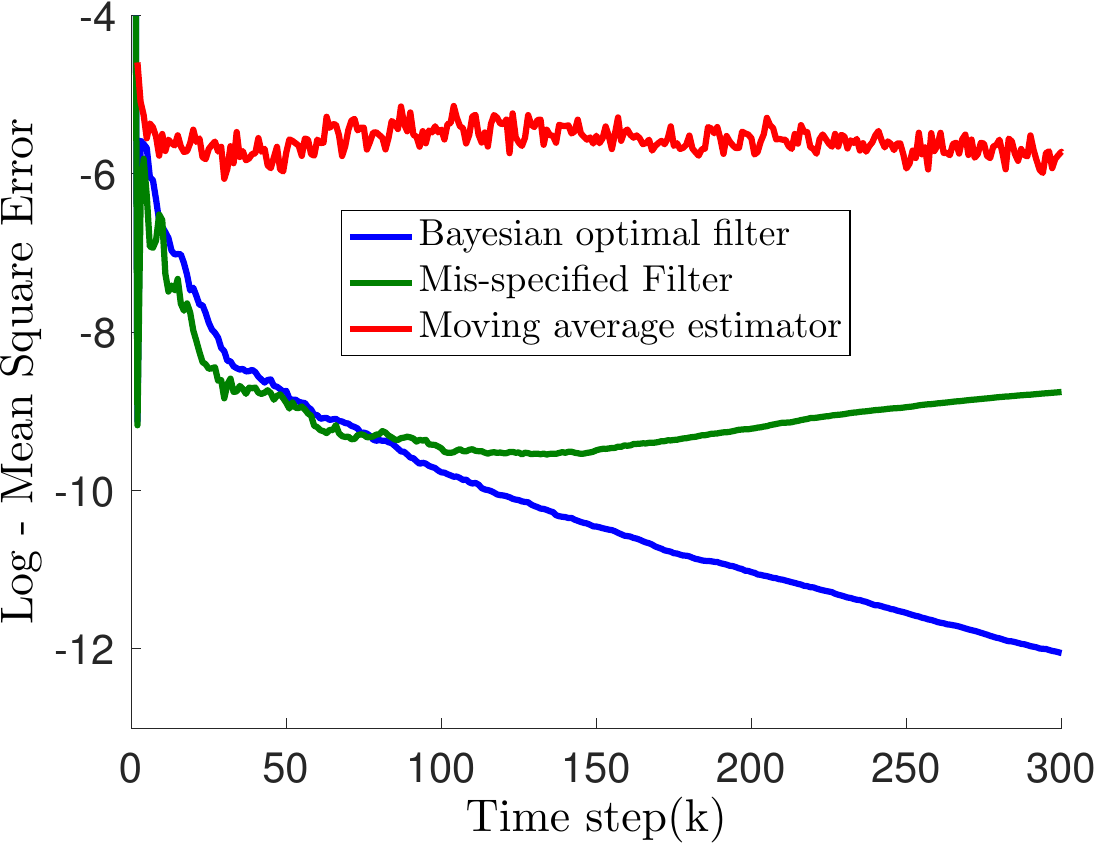}
	\caption{Comparison of the mean square error between a Bayesian filter, a mis-specified Bayesian filter and a vector autoregressive estimator. }
	\label{fig:MSE_incorrectFilter2}
\end{figure}

 We call a degree distribution mis-specified if it does not match the degree distribution of the true network. A Bayesian filter, derived with a scale-free degree distribution; a mis-specified filter, same Bayesian filter derived with the degree distribution $\degdist(l) = \frac{1}{L}   \quad \forall l$; and an autoregressive moving average filter with parameters computed by multivariate least square estimation (for a comprehensive review of vector auto regression and LS parameter estimation see \cite{Ham94}) are used to analyze the sensitivity of filter performance to mis-specified models. It is observed in Fig.~\ref{fig:MSE_incorrectFilter2} that, even when the degree distribution is mis-specified, the Bayesian filters outperform the moving average filter with an MSE of the order of $10^{-8}$, compared to $10^{-6}$ of the moving average filter.

\subsection{Analysis and Validation on Twitter Dataset}

This section illustrates the tracking of infection diffusion on social networks using real diffusion data from the microblog platform Twitter. Twitter is a social media platform over which users can communicate in short $<140$ character messages, images and video files.  We analyze the diffusion of information through the Twitter Social network to demonstrate the effectiveness of the SIS model of Sec.~\ref{sec:modeldef}. \\ 
Twitter played a critical role in the Egyptian revolution of $2011$ or January $25^{th}~(\#\text{Jan}25)$ uprising \cite{PO12}. Twitter was used by protesters to organize the protest and recruit members and as a medium to discuss and share information about the protest. This uprising precipitated quickly and was violent, both of which acted as substantial barriers towards traditional media coverage. Below, we refer to the interest and engagement with the news of the uprising as \textit{infection} and track the distribution of infection over time. 

\subsubsection{Dataset} The dataset consists of tweets sampled between January~$23^{rd}$ and February~$8^{th},~2011$ and are available from Twitter (http://trec.nist.gov/data/tweets/). The tweet collection period encapsulates the time-frame of the first major developments relating to the January~$25^{th}$ uprising event. In Twitter, a ``hashtag" follows the discussion topics, i.e., a word or a phrase prefixed with the number sign \#. We make use of the hashtags to track the spreading of a specific topic on Twitter. The most used hashtag related to this protest is ``\#Jan25"€. To obtain the information spreading among users participating in this protest, we filtered 26,313 tweets containing ``\#Jan25" published by 13,341 different users, from around $10$ million tweets. These tweets contain the event of interest and the social network is (re-)constructed from them as follows: two users are connected if one user has mentioned another user (``@username") in the tweet containing ``\#Jan25" at least once over the duration of interest. On this constructed social network, information diffusion is analysed. \\
All users in the constructed social network are assumed to be susceptible initially. Users who initiate tweets on the event of interest are assumed to be infected and act as seeds for the spread of information. Once a user, say User\#A becomes infected, it has some constant probability, say $\decay$, of becoming susceptible in each time period. This modeling assumption is motivated by the frequently observed poisson-like decay of an individual's interest in social media topics  \cite{CS08}. 


\subsubsection{SIS model for Twitter data}

 In case of the spread of engagement in Twitter, individuals can either be inactive or active depending on if they are willing and able to spread information and interest on a topic. Active users can be considered `infected', and inactive users can become `infected' by interacting with other `infected' individuals, in particular, any of its active neighbours in a social network. In this way, engagement and knowledge of a topic spreads throughout the network. People can also become disinterested in a subject they have already been exposed to, in this way they are not currently engaged, but may become engaged if contacted by an infected neighbour; thus inactive individuals are assumed to be susceptible. This is the basis of the susceptible-infected-susceptible (SIS) model and its application to the diffusion of information on Twitter has been noted and studied; see~\cite{KLPM10},~\cite{JDSCR13}.
 
 We adopt the SIS model for the Twitter data, mapping engagement in the January~$25^{th}$ uprising as an infection spreading over the Twitter network created from Twitter users. Below we analyze the goodness-of-fit of the SIS model for the Twitter data using a standard statistical test, the Kolmogorov-Smirnov test \cite{Mas51}, and measure the square and absolute difference between the data and model predictions. The SIS model used here for evaluation utilizes empirically determined parameters as outlined in Sec.~\ref{sec:twitsample}.\\ 
\textit{Model Evaluation (Goodness of Fit for SIS model)}: We used the Kolmogorov-Smirnov (KS) test on the empirical infected degree distribution at the final timepoint to evaluate the SIS model. The KS test statistic was 0.2286 with p-value 0.2813. The null hypothesis for this statistical test is that both the observed Twitter and SIS model infected degree distributions are samples of the same infected degree distribution. At a confidence level of 0.01, we do not reject the null hypothesis. We also calculate the average and maximum square difference between the Twitter data and predicted SIS degree infection probabilities. The differences are calculated at each timepoint; both the averaged-over-time and maximum differences are shown. These values can be seen in Table~\ref{tb:deviation} and the trajectories are shown in Fig.~\ref{fig:TwitSim}.
\begin{table}[ht]
\begin{center}
    \caption{Goodness of fit for the SIS model: The average and maximum deviations between the Twitter data and SIS model predictions are presented. The large absolute difference in high degree nodes arises when there are few nodes of this large degree and thus the empirical infection probability deviates from the asymptotic case.}
    \begin{tabular}{ |p{2.9cm}  | p{1.1cm}  | p{1.1cm} | p{1.1cm} |p{1.1cm} |}
    \hline
    Degree &1  & 2 & 3+ \\  \hline
    \raggedright Average Square Difference   & 0.0011 & 0.0014 & 0.0235 \\ \hline
    \raggedright Average Absolute Difference  & 0.0273 & 0.0294 & 0.1000 \\ \hline 
    \raggedright Maximum Absolute Difference  & 0.0644 & 0.0719 & 0.8403 \\ \hline 
    \end{tabular}  \label{tb:deviation}
\end{center}
\end{table}\\
The low magnitude of the model deviations in Table~\ref{tb:deviation} for the Twitter dataset and the failure to reject the hypothesis that the Twitter data and model infected degree distributions come from the same distribution, suggest that the SIS model is a satisfactory model with respect to the infection dynamics of interest in the January~$25^{th}$ uprising.
\subsubsection{Sampling for tracking the infected degree distribution}
\label{sec:twitsample}
\begin{figure}
     \centering
     \includegraphics[width=3.4in]{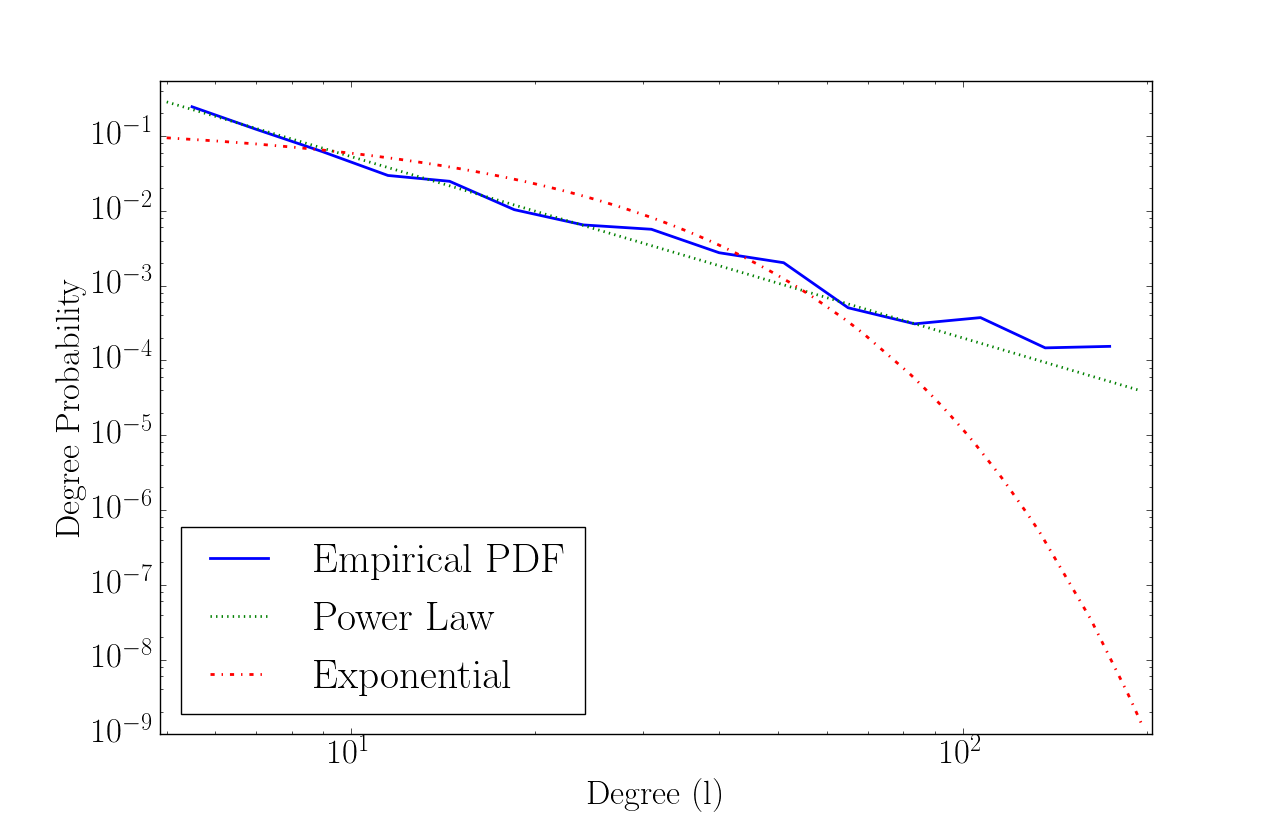}
     \caption{The degree distribution of the constructed Twitter \#Jan25 social network is analyzed and found to follow a power law distribution. The exponent parameter  $-2.425$ of the power law distribution was chosen as the maximum likelihood estimate. The MLE and likelihood ratio were computed by the computational package in \cite{ABP14} according to the methods in \cite{CSN09}. The loglikelihood ratio between a power law and exponential distributions is $8.631$, which is significant evidence that the data follows a power law distribution rather than an exponential distribution. }
     \label{fig:TwitDegree}
 \end{figure}

\begin{figure}
     \centering
     \includegraphics[width=2.6in]{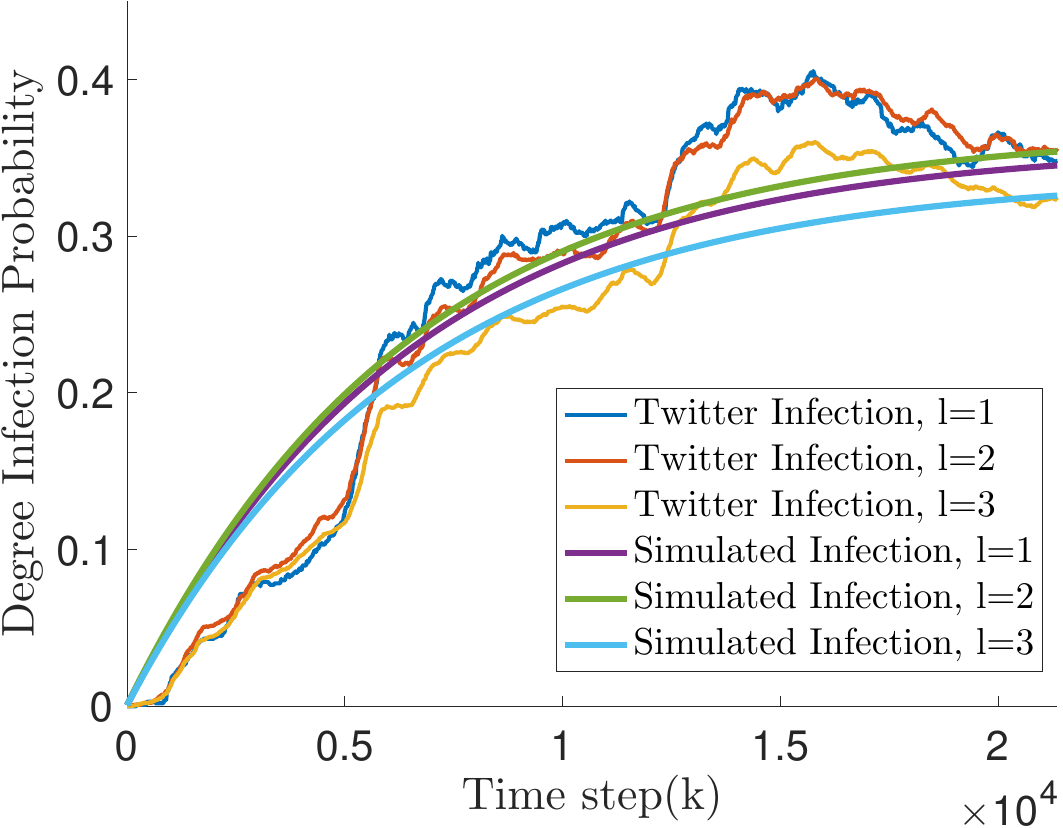}
     \caption{The true Twitter infection and simulated Twitter infections are compared for nodes of degree $\l = 0, 1, 2$. The simulated trajectories use the empirically generated $\degdist$, $\decay$ and $\hat{\tp}_{12}$ (\ref{eq:emp_tr}).}
     \label{fig:TwitSim}
 \end{figure}

The mean field dynamics for the SIS model can be used to track and predict the evolution of the infection on Twitter. We must generate estimates of $\tp_{12}$, $\tp_{21}$, and determine the degree distribution from samples obtained from (\ref{eq:obs_inf}). $\tp_{12}$ is given by $\decay$, since all infected nodes become susceptible with probability $\decay$ at each time point. We compute the empirical transmission rates $\hat{\tp}_{21}$ directly by observing the frequency with which an infected individual with $\l$ neighbors, $a$ of which are infected, becomes infected. 
\begin{equation} \label{eq:emp_tr}
\hat{\tp}_{12}(l,a) = \frac{\sum\limits_{n=0}^{T} \sum\limits_{m=1}^{M} \left(s^{(m)}_{n+1} = 1 | s^{(m)}_n = 2, D^{(m)}=l, F_n^{(m)} = a \right)}  {\sum\limits_{n=0}^{T} \sum\limits_{m=1}^{M} \left( s^{(m)}_n = 2, D^{(m)}=l, F_n^{(m)} = a \right)} 
 \end{equation}
The degree distribution used is the empirical degree distribution, shown in Fig.~\ref{fig:TwitDegree}. We analyze the degree distribution and find that it fits a power law distribution with power law exponent $-2.425$ which matches very closely with the power law distribution with exponent$-2.412$ found for the Twitter network in \cite{JSFT07}. The true degree infection probabilities are computed directly from the entire network for each 1 minute time interval.

\begin{figure}
     \centering
     \includegraphics[width=2.6in]{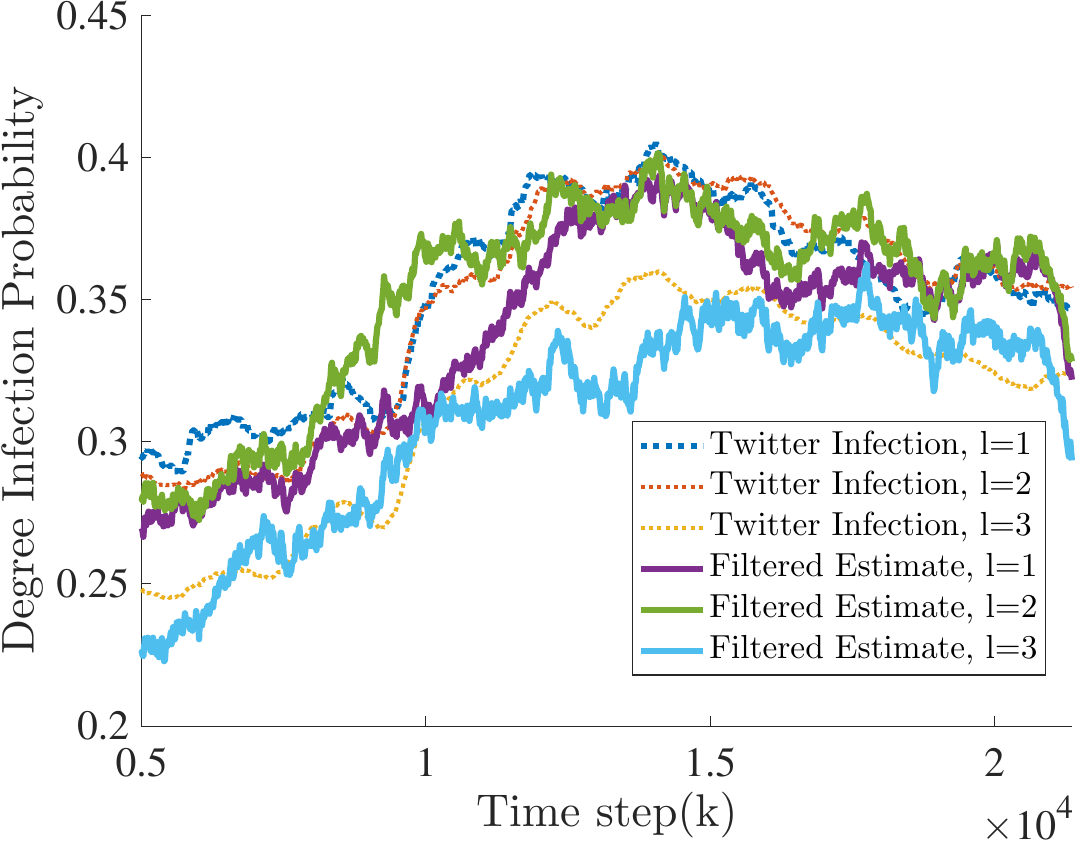}
     \caption{The true Twitter infection and the filter estimates of the Twitter degree infection probabilities are compared. Samples are generated by RDS sampling on the \#Jan25 social network. At each timestep a 10,000 node walk is performed and from this walk an observation of the infected degree distribution is generated. }
     \label{fig:TwitFilter}
 \end{figure}

Next, we sample only a subset of the data using the RDS sampling scheme described in Sec.~\ref{subsec:samp}, every $1$~minute and track the infected degree distribution over time using the non-linear Bayesian Filtering technique described in Sec.~\ref{sec:filter}. The parameters used in the Bayesian filter are: empirical $\hat{\tp}_{12}$ and empirical $\rho$ and the filter estimates are shown in Fig.~\ref{fig:TwitFilter}. It is seen that the filtered estimates track the true infected degree distribution satisfactorily over time.

\section{Conclusion}

We considered the problem of tracking infection diffusion over large social networks by modeling the diffusion process using SIS model. The infection distribution was approximated using mean field dynamics which resulted in a state space model with polynomial dynamics. Posterior Cram\'er-Rao bounds were computed for the mean field dynamics and it was shown that these bounds are relatively insensitive to the type of underlying social network (Erd\H{o}s-R\'enyi vs Scale Free network). Next, to account for the time-varying nature of the degree distribution of large real-world networks, the relationship between diffusion thresholds and the changing degree distribution was analyzed using a generative model for a network generated using the preferential attachment model. It was found that networks on which more edges are added relative to nodes, have lower diffusion thresholds. Finally, a Twitter dataset was used to illustrate how information diffusion on the Twitter platform can be modeled by a mean field dynamical SIS model, and that, under this model, we can filter and track the evolution of the degree infection probabilities over time.

\appendices

\section{Proof of Mean Field Theorem}
The proof of the mean field dynamics approximation is given in \cite{BW03}, but the presentation is not readily accessible. We show below that the proof is a simple consequence of Azuma-Hoeffding inequality. A bound on the deviation between the mean field dynamics $\bst_n$ in (\ref{eq:mfd}) and actual population distribution $\stn$ in (\ref{eq:mrmk}) is calculated in the form of two lemmas, Lemma~\ref{lem:mftstep1} and Lemma~\ref{lem:mftstep2}. We first state the Azuma-Hoeffding Inequality (Theorem~\ref{thm:az_hf}~below) which gives a bound on the deviation of a random variable from some value for the values of martingales that have bounded differences. 

\begin{theorem}[Azuma-Hoeffding Inequality] \label{thm:az_hf}
Suppose $S_\finaltime = \sum_{k=1}^\finaltime v_{k} + S_0$ where $\{v_k\}$ is a martingale difference process with bounded differences satisfying $|v_k|\leq \Delta_k$ almost surely where $\Delta_k$ are finite constants. Then for any $\epsilon >0$,
$$
\prob (|S_\finaltime - S_0| \geq \epsilon) \leq 2 \exp \left(- \frac{\epsilon^2}{\sum_{k=1}^\finaltime \Delta_k^2} \right)
$$
\qed
\end{theorem}

Define
$$ \dev_n = \popn - \bpopn, \quad  S_\finaltime = \max_{1 \leq n \leq \finaltime} \| \sum_{k=1}^n \mtgk\|_\infty$$ Here, $v_k$ is an $L$-dimensional finite-state martingale increment process with $\|v_k\|_2 \leq \frac{\Gamma}{\vertexnum}$ for some positive constant $\Gamma$. 

\begin{lemma} \label{lem:mftstep1} 
$$ \|\dev_{n+1}\|_\infty  \leq \|\dev_0 \|_\infty + \frac{\beta}{M} \sum_{k=1}^n \| \dev_k \|_\infty + S_\finaltime .$$
\end{lemma}

{ \textbf{Proof of Lemma \ref{lem:mftstep1}}}: Recall
$ \dev_n = \stn -  \bst_n $. Let $H(x)$ be a Lipschitz function. Clearly,
\begin{align*}
\dev_{n+1}  &= \dev_n + \frac{1}{M} [H(\stn) - H(\bst_n)] + v_n \\
&= \dev_0 + \frac{1}{M} \sum_{k=1}^n [H(\popn) - H(\bpopn)] + \sum_{k=1}^n \mtgk \\
\|\dev_{n+1}\|_\infty & \leq \|\dev_{0}\|_\infty + \frac{1}{M} \sum_{k=1}^n\| [H(\popn) - H(\bpopn)]\|_\infty +  \|\sum_{k=1}^n \mtgk\|_\infty \\
& \leq
 \|\dev_{0}\|_\infty + \frac{\beta}{M} \sum_{k=1}^n\| \dev_k\|_\infty + S_\finaltime
\end{align*} 
since $\| [H(\popn) - H(\bpopn)]\|_\infty \leq \beta \|\popn - \bpopn\|_\infty$ where $\beta$ is the Lipschitz constant. \qed

\begin{lemma}\label{lem:mftstep2}  
$$ \prob\bigl(S_\finaltime \geq \epsilon \bigr) \leq 2 \exp\left( - \frac{\epsilon^2 M^2}{2 \Gamma \finaltime} \right)$$
\end{lemma}

{ \textbf{Proof of Lemma \ref{lem:mftstep2}}}:
$ \| \sum_{k=1}^n \mtgk \|_\infty  = \max_i  | \sum_{k=1}^n e_i^\p \mtgk |  =  | \sum_{k=1}^n e_{i^*}^\p \mtgk | $
for some $i^*$.
Since $e_{i^*}^\p \mtgk$ is  a martingale difference process with $|e_{i^*}^\p \mtgk| \leq \sqrt{\Gamma}/M$ applying the Azuma-Hoeffding inequality (Theorem~\ref{thm:az_hf}) yields   
$$ \prob(\| \sum_{k=1}^n \mtgk \|_\infty \geq \epsilon) = \prob(| \sum_{k=1}^n e_{i^*}^\p \mtgk | \geq \epsilon) \leq
2 \, \exp \left[ - \frac{\epsilon^2 M^2}{2 \Gamma n} \right]
$$
The right hand side is increasing with $n$. So clearly
$$  \prob(
\max_{1 \leq n \leq \finaltime} \| \sum_{k=1}^n \mtgk \|_\infty \geq \epsilon) \leq
2 \, \exp \left[ - \frac{\epsilon^2 M^2}{2 \Gamma \finaltime } \right] $$ \qed


Using Lemmas \ref{lem:mftstep1} and  \ref{lem:mftstep2} the proof of Theorem \ref{thm:mfd} is as follows.
Applying Gronwall's inequality\footnote{Gronwall's inequality: if $\{x_k\}$ and $\{b_k\}$ are non-negative sequences
and $a \geq 0$, then $$\stn \leq a + \sum_{k=1}^{n-1} x_k b_k
\implies
 x_n \leq a \exp(\sum_{k=1}^{n-1} b_k ) $$ }
 to Lemma \ref{lem:mftstep1} yields $ \|\dev_n\|_\infty  \leq  S_\finaltime \exp\left[ \frac{  \beta n}{M} \right] $, which in turn implies that
 $$ \max_{1 \leq n \leq \finaltime}  \|\dev_n\|_\infty \leq  S_\finaltime \exp\left[ \frac{  \beta \finaltime}{M} \right] .$$
As a result
\begin{align*}
\prob( \max_{1 \leq n \leq \finaltime}  \|\dev_n\|_\infty  > \epsilon) &\leq \prob(S_\finaltime  \exp\left[ \frac{  \beta \finaltime}{M} \right] > \epsilon) \\
&=  \prob\bigl(S_\finaltime >   \exp\left[- \frac{  \beta \finaltime}{M} \right] \, \epsilon\bigr)
\end{align*}
Next applying Lemma \ref{lem:mftstep2} to the right hand side yields
$$ \prob( \max_{1 \leq n \leq \finaltime}  \|\dev_n\|_\infty  > \epsilon) \leq
2 \exp\left( - \exp\bigl(\frac{-2 \beta \finaltime}{M} \bigr)\, \epsilon^2  \frac{M^2}{2 \Gamma \finaltime} \right) .$$
Finally choosing $\finaltime = c_1 M$, for some positive constant $c_1$ yields
\begin{align*}
\prob( \max_{1 \leq n \leq \finaltime}  \|\dev_n\|_\infty  > \epsilon) \leq 2 \exp( - C_2 \epsilon^2 M)
\end{align*}
where $C_2 = \exp( - 2 \beta c_1) \frac{1}{2 \Gamma c_1}$. This completes the proof of Theorem \ref{thm:mfd}. \qed


\section{Recursion for Posterior Cramer Rao Lower Bound \cite{TMN98}}
Consider a non-linear state space model given by:
\begin{equation*}
\begin{split}
\stnp = f(\stn) + w_n \\
\obs_{n} = h(\stn) + v_n 
\end{split} 
\end{equation*}
with $w_n \sim \mathscr{N}(0,\mathbf{Q}_n)$ and $v_n \sim \mathscr{N}(0,\mathbf{R}_n)$.
Then, the recursive equations for parameters estimation are given by \cite{TMN98}: 
$$
J_{n+1} = D_n^{22} - D_n^{21} \left( J_n +D_n^{11}\right)^{-1} D_n^{12}
$$
where 
\begin{equation} \label{eq:pclrb}
\begin{split}
D_n^{11} &=  \mathbb{E} \{ -\Delta_{\stn}^{\stn} \log p(\stnp |\stn) \} \\
D_n^{12} &=  \mathbb{E} \{ -\Delta_{\stn}^{\stnp} \log p(\stnp|\stn) \} \\
D_n^{21} &=  \mathbb{E} \{ -\Delta_{\stnp}^{\stn} \log p(\stnp |\stn) \} \\
D_n^{22} &=  \mathbb{E} \{ -\Delta_{\stnp}^{\stnp} \log p(\stnp |\stn) \} \\
\quad & \quad + \mathbb{E} \{ -\Delta_{\stnp}^{\stnp} \log p(\obs_{n+1} |\stnp) \}
\end{split}
\end{equation}
As was stated in Sec.~\ref{sec:modeldef}, the state evolution has polynomial dynamics $f(\cdot)$ and sampling results in linear observations $C$. Since our model has Gaussian state (\ref{eq:per_st}) and observation noise (\ref{eq:obs_inf}), we can compute the components of (\ref{eq:pclrb}) in terms of the state and observation functions $f(\cdot)$ and $C$. 
\begin{equation}
\begin{aligned}
{\hspace{-2cm}}-\log p(\stnp|\stn) &=\\  c + \frac{1}{2}\{ \stnp - &f_n(\stn) \}^\prime \textbf{Q}_n^{-1} \{\stnp -f_n(\stn)\} \\
{\hspace{-1.8cm}}-\log p(\obs_{n+1}|\stnp) &=\\ c + \frac{1}{2}\{ \obs_{n+1} -&C \stnp \}^\prime \textbf{R}_n^{-1} \{ \obs_{n+1} -C \stnp\}
\end{aligned}
\end{equation}
Thus $D_n^{11}, D_n^{21},D_n^{12},D_n^{22}$ are given as:
\begin{equation}
\begin{split}
 D_n^{11} &=  \mathbb{E} \{ (\Delta_{\stn}f_n^\prime(\stn))\textbf{Q}_n^{-1} (\Delta_{\stn}f_n^\prime(\stn))^\prime \} \\
 D_n^{12} &=   \mathbb{E} \{ \Delta_{\stn}f_n^\prime(\stn) \} \textbf{Q}_n^{-1}  \\
 D_n^{21} &=  \{D_n^{12}\}^\prime \\
 D_n^{22} &=  \textbf{Q}_n^{-1} +C \textbf{R}_n^{-1} C^\prime  \\
\end{split}
\end{equation}

 \section{Proof of Theorem~\ref{thm:cmp_gr}}
 \subsection{Definitions}
Let $\Pi(X) {\overset{\Delta}{=}} \{ \degdist\in\mathbb{R}^{X} : \textbf{1}_{X}'\degdist = 1, 0\leq \degdist(i) \leq 1 ~ \text{for all} ~ i\in \{ 1,2, \hdots, X \} \}$ denote the $X-1$ dimensional unit simplex. 
 \begin{definition}{\textit{First-Order Stochastic Dominance} ($\geq_{sd}$):} Let $\degdist_{1},\degdist_{2}\in \Pi(X)$ be any two belief state vectors. Then $\degdist_{1}\geq_{sd}\degdist_{2}$ if 
\begin{equation*}
{\overset{X}{\underset{i=j}{\sum}}} \degdist_{1}(i) \geq {\overset{X}{\underset{i=j}{\sum}}} \degdist_{2}(i) ~ \text{for} ~ j \in \{1,\hdots,X\}.
\end{equation*}  
\end{definition}
 \begin{definition}{\textit{Second-Order Stochastic Dominance} ($\geq_{ssd}$):} Let $\degdist_{1},\degdist_{2}\in \Pi(X)$ be any two belief state vectors with $F_1$ and $F_2$ as the corresponding cumulative distribution functions. Then $\degdist_{1}\geq_{ssd}\degdist_{2}$ if 
\begin{equation*}
{\overset{i}{\underset{j=1}{\sum}}} F_{1}(i) \leq {\overset{i}{\underset{j=1}{\sum}}} F_{2}(i) ~ \text{for} ~ i \in \{1,\hdots,X\}.
\end{equation*}  
\end{definition}

\begin{lemma}{\normalfont{\cite{MS02}}} \label{lem:dec}
$\degdist_{2} \geq_{s} \degdist_{1}$ iff for all $v\in \mathcal{V}$, $v'\degdist_{2}\leq v'\degdist_{1}$, where $\mathcal{V}$ denotes the space of $X$- dimensional vectors $v$, with non-increasing components, i.e, $v_{1} \geq v_{2} \geq \hdots v_{X}$.
\end{lemma}
\begin{lemma}{\normalfont{\cite{MS02}}} \label{lem:inc}
$\degdist_{2} \geq_{s} \degdist_{1}$ iff for all $v\in \mathcal{V}$, $v'\degdist_{2}\geq v'\degdist_{1}$, where $\mathcal{V}$ denotes the space of $X$- dimensional vectors $v$, with non-decreasing components, i.e, $v_{1} \leq v_{2} \leq \hdots v_{X}$.
\end{lemma}

\subsection{Proofs}
\begin{theorem}[\cite{Pin08}] \label{thm:df_thr}
Any two networks having degree distributions $\degdist^1$ and $\degdist^2$ respectively, with $\degdist^1 <_{ssd}\degdist^2$, the diffusion threshold $\difth^1 > \difth^2$. \footnote{ $<_{ssd}$ denotes second order stochastic dominance. } \qed
\end{theorem}
In words, as the number of nodes with higher degree increase, the probability of a large fraction of agents becoming infected increases.
\begin{lemma} \label{lem:st_idm}
For any $p\in [0,1]$, the transition matrix $H_\dtme(p)$ for a preferential attachment graph is such that $$H_\dtme^{i}(p) <_{sd} H_\dtme^{i+1}(p)$$ for $i=1,2,\hdots$, where $H_\dtme^{i}(p)$ denotes the $i^{th}$ row of $H_\dtme(p)$.
\end{lemma}
\textbf{Proof of Lemma~\ref{lem:st_idm}}:
It is clear from the definition of $\degtr_{\dtme}(p)$ that each row has only 2 non-zero elements: probability of node undergoing no change and probability of having a degree lesser by 1 and an adjacent edge was added during evolution~(\ref{eq:deg_ev}). From the definition of First-order stochastic dominance, we have that a row dominates another row if the tail sum of the former is larger than the latter. Since the matrix $\degtr_{\dtme}(p)$ is upper bidiagonal, the result follows. \qed 
\begin{lemma} \label{lem:st_bdm}
Let $H_\dtme(p_{1})$ and $H_\dtme(p_{2})$ be two social networks modeled using preferential attachment with the probability of adding a new vertex, $p_i>0$.  If $p_1>p_2$, then\footnote {${\overset{r}>}_{sd}$  denotes row-wise first order stochastic dominance. $H^i_\dtme(p_{2}) ~>_{sd}~H^i_\dtme(p_{1})$ for $i = 1,2,\hdots$.} $$H_\dtme(p_{2}) {\overset{r}>}_{sd} H_\dtme(p_{1}) $$
\end{lemma}
\textbf{Proof of Lemma~\ref{lem:st_bdm}}:
Given $p_1>0$, $p_2>0$ and $p_1 > p_2$. Since First-order dominance is the comparison of tail sums and since each row has only 2 non-zero elements, to compare the same rows of 2 different matrices, it is sufficient to compare the first element. For the sake of illustration, let us consider row $i$. Since $i>0$ and $\dtme>0$,
\hspace{-2cm}$$p_1>p_2 \Rightarrow 2-p_1<2-p_2 \Rightarrow i(2-p_1)<i(2-p_2)$$
$$\Rightarrow \frac{i(2-p_1)}{\dtme} < \frac{i(2-p_2)}{\dtme} \Rightarrow 1-\frac{i(2-p_1)}{\dtme} > 1-\frac{i(2-p_2)}{\dtme}$$ and therefore $H_\dtme(p_{2}) {\overset{r}>}_{sd}  H_\dtme(p_{1})$. \qed

Lemma~\ref{lem:st_idm} says that the rows of the transition matrix are first-order increasing. Lemma~\ref{lem:st_bdm} says that the transition probabilities are first-order increasing in the probability of adding new edges, $1-p$. 
\begin{lemma} \label{lem:tr_suf1}
Let $H_\dtme(p)$ be such that $H_\dtme^{i}(p) <_{sd} H_\dtme^{i+1}(p)$ for $i=1,2,\hdots$, where $H_\dtme^{i}(p)$ denotes the $i^{th}$ row of $H_\dtme(p)$. Then for any probability distributions $\degdist_1$ and $\degdist_2$ with $\degdist_1 <_{sd} \degdist_2$, $$ H_\dtme^\prime(p) \degdist_1 <_{sd} H_\dtme^\prime(p) \degdist_2$$
\end{lemma}
\textbf{Proof of Lemma~\ref{lem:tr_suf1}}:
For convenience, let $\Psi=H_\dtme^\prime(p_{2})$. From the definition of First-order dominance on the last row $$\Psi_{1N^+} \leq \Psi_{2N^+} \leq \Psi_{3N^+} \hdots \Psi_{N^+N^+} $$From Lemma~\ref{lem:inc}, we have $$\sum_i \degdist_1^i \Psi_{iN^+} \leq \sum_i \degdist_2^i \Psi_{iN^+} $$
Any (arbitrary) $s^{th}$ element for the two distribution vectors is given by $$\sum_i \degdist_1^i \sum_{k=s}^{N^+} \Psi_{ik}~~\text{and}~~\sum_i \degdist_2^i \sum_{k=s}^{N^+} \Psi_{ik}$$
We have from the definition of First-order dominance, $$\sum_{k=s}^{N^+} \Psi_{ik} \leq \sum_{k=s}^{N^+} \Psi_{(i+1)k}\quad \text{for}~i \in \{1,2\hdots,N\}$$
From Lemma~\ref{lem:inc} and using $\degdist^1<_{sd}\degdist^2$, 
\begin{align*}
\sum_i \degdist_1^i \sum_{k=s}^{N^+} \Psi_{ik} &\leq \sum_i \degdist_2^i \sum_{k=s}^{N^+} \Psi_{ik} \\
\sum_{k=s}^{N^+} \sum_i \degdist_1^i \Psi_{ik} &\leq \sum_{k=s}^{N^+} \sum_i \degdist_2^i \Psi_{ik} \\
\Psi \degdist_1 &<_{sd} \Psi \degdist_2
\end{align*}
where $\degdist^i_{*}$ denotes the $i^{th}$ element of the distribution vector~$\degdist_{*}$.
\qed
\begin{lemma} \label{lem:tr_suf2}
Let $p_1>p_2$ and $H_\dtme(p_{2}) {\overset{r}>}_{sd} H_\dtme(p_{1})$. Then for any probability distribution $\degdist$, $$H_\dtme^\prime(p_{1}) \degdist <_{sd}H_\dtme^\prime(p_{2}) \degdist$$
\end{lemma}
\textbf{Proof of Lemma~\ref{lem:tr_suf2}}:
For convenience, let $\Psi^2=H_\dtme^\prime(p_{2})$ and $\Psi^1=H_\dtme^\prime(p_{1})$. We know that the maximum degree is represented as $N^+$. From the definition of First-order dominance on the last row, 
\begin{align*}
\Psi^1_{iN^{+}} &\leq \Psi^2_{iN^{+}}~\text{for all $i$}. \\
\Rightarrow \sum_{i} \degdist^i \Psi^1_{iN^{+}} &\leq \sum_{i} \degdist^i \Psi^2_{iN^{+}}
\end{align*}

Let $s \in \{1,2,\hdots,N\}$ be arbitrary. From the definition of First-order stochastic dominance on corresponding rows of $\Psi^1$ and $\Psi^2$,
\begin{align*}
\sum_{k=s} \Psi^1_{ik} &\leq \sum_{k=s} \Psi^2_{ik}~\text{for all $i$}. \\
\Rightarrow \sum_{i} \degdist^i \sum_{k=s}^{N^+} \Psi^1_{ik} &\leq \sum_{i} \degdist^i \sum_{k=s}^{N^+} \Psi^2_{ik} \\
\Rightarrow \sum_{k=s}^{N^+} \sum_{i} \degdist^i \Psi^1_{ik} &\leq \sum_{k=s}^{N^+} \sum_{i} \degdist^i \Psi^2_{ik} \\
\Rightarrow \Psi^1 \degdist &<_{sd} \Psi^2 \degdist
\end{align*} 
where $\degdist^i$ denotes the $i^{th}$ element of the distribution vector~$\degdist$. 
\qed

\textbf{Proof of Theorem~\ref{thm:cmp_gr}}:
\begin{enumerate}
\item The proof of Theorem~\ref{thm:cmp_gr}.$1$ is by induction on $\dtme$. Consider an initial probability distribution $\degdist_0$. \\
\textit{Base}: Let $p_1>p_2>0$. From Lemma~\ref{lem:tr_suf2},
$$
\degdist_1(p_2)=H_\dtme^\prime(p_{2}) \degdist_0 >_{sd}\degdist_1(p_1)=H_\dtme^\prime(p_{1}) \degdist_0  
$$ 
 By Lemma~\ref{lem:tr_suf1} and Lemma~\ref{lem:tr_suf2},
\begin{align*}
H_\dtme^\prime(p_1) \degdist_{1}(p_1) &<_{sd} H_\dtme^\prime(p_1)\degdist_{1}(p_2) \\
H_\dtme^\prime(p_1)\degdist_{1}(p_2) &<_{sd} H_\dtme^\prime(p_2)\degdist_{1}(p_2)  \\
\Rightarrow H_\dtme^\prime(p_1) \degdist_{1}(p_1) &<_{sd} H_\dtme^\prime(p_2)\degdist_{1}(p_2) \\
\Rightarrow (H_\dtme^\prime(p_{1}))^2 \degdist_0 &<_{sd}(H_\dtme^\prime(p_{2}))^2 \degdist_0  
\end{align*}
\textit{Induction step}: Let the result hold for all $\dtme \leq q$. 
$$
(H_\dtme^\prime(p_{2}))^q \degdist_0 >_{sd}(H_\dtme^\prime(p_{1}))^q \degdist_0  \\
$$ 
Let $(H_\dtme^\prime(p_{2}))^{q} \degdist_0 = \degdist_{q+1}(p_2)$ and $(H_\dtme^\prime(p_{1}))^{q} \degdist_0 = \degdist_{q+1}(p_1)$. By Lemma~\ref{lem:tr_suf1} and Lemma~\ref{lem:tr_suf2},
\begin{align*}
H_\dtme^\prime(p_1) \degdist_{q+1}(p_1) &<_{sd} H_\dtme^\prime(p_1)\degdist_{q+1}(p_2) \\
H_\dtme^\prime(p_1)\degdist_{q+1}(p_2)&<_{sd} H_\dtme^\prime(p_2)\degdist_{q+1}(p_2)  \\
\Rightarrow H_\dtme^\prime(p_1) \degdist_{q+1}(p_1) &<_{sd} H_\dtme^\prime(p_2)\degdist_{q+1}(p_2)\\
\Rightarrow (H_\dtme^\prime(p_{1}))^{q+1} \degdist_0 &<_{sd}(H_\dtme^\prime(p_{2}))^{q+1} \degdist_0  
\end{align*}
As $q$ is any arbitrary positive integer, the result holds for all $\dtme>1$. \qed
\item The proof of Theorem~\ref{thm:cmp_gr}.$2$ easily follows from Theorem~\ref{thm:cmp_gr}.$1$ and Theorem~\ref{thm:df_thr} and is omitted. Using Theorem~\ref{thm:cmp_gr}, the degree distributions are ordered, first order stochastic dominance implies second order dominance \cite{Kri16}, and from Theorem~\ref{thm:df_thr}, the result follows. \qed
\end{enumerate}

\ifCLASSOPTIONcaptionsoff
  \newpage
\fi



\bibliographystyle{IEEEtran}
%
\bibliography{references}

%








\end{document}